\documentclass[12pt]{article}
\usepackage{a4wide,feynmf,amsmath,amssymb,graphicx,psfrag}

\newcommand{\ehsp}{\hspace{0.25cm}}

\newcommand{\dvk}{d^4\!k}
\newcommand{\idvk}{\int\!\!\dvk}

\newcommand{\polv}[3]{\varepsilon_{#1}^{#2}(#3)}

\newcommand{\gev}{\text{GeV}}

\newcommand{\fb}{\text{fb}}

\newcommand{\MM}{{\cal M}}
\newcommand{\Sq}{   {\tilde{q}} }
\newcommand{\Stop}{ {\tilde{t}} }
\newcommand{\Sbot}{ {\tilde{b}} }

\newcommand{\mhpm}{ m_{H^{\pm}} }

\newcommand{\MSQ}{ M_{\tilde{Q}} }
\newcommand{\MSU}{ M_{\tilde{U}} }
\newcommand{\MSD}{ M_{\tilde{D}} }

\newcommand{\sqrts}{\sqrt{\hat{s}}} 

\begin{document}
\setlength{\unitlength}{1mm}


\begin{titlepage}
\begin{flushright}
{\bf KA--TP--14--2000\\
hep-ph/0008308}
\end{flushright}
\vspace{2cm}
\begin{center}
{\Large \bf The MSSM prediction for $W^\pm H^\mp$ production \\[0.4cm]
  by gluon fusion 
at the Large Hadron Collider} \\[2.5cm]
{\large \bf Oliver~Brein \footnote{E-mail: obr@itp.uni-karlsruhe.de},
Wolfgang~Hollik \footnote{E-mail: Wolfgang.Hollik@physik.uni-karlsruhe.de}, 
Shinya~Kanemura \footnote{E-mail: kanemu@itp.uni-karlsruhe.de
}}  \\[0.8cm]
{\normalsize\em 
        Institut f\"ur Theoretische Physik, Universit\"at Karlsruhe,\\
        D-76128 Karlsruhe, Germany}
\end{center}
\vspace{2cm}
\begin{abstract}  
We discuss the associated $W^\pm H^\mp$ production in $pp$ collision 
for the Large Hadron Collider. A complete one-loop calculation of the 
loop-induced subprocess $gg \to W^\pm H^\mp$ is presented in the framework 
of the Minimal Supersymmetric Standard Model (MSSM), and the possible
enhancement of the hadronic cross section is investigated under the
constraint from the squark direct-search results and the low-energy
precision data. Because of the large destructive interplay in the
quark-loop contributions between triangle-type and box-type diagrams,
the squark-loop contributions turn out to be comparable with the
quark-loop ones.
In particular, the hadronic cross section via gluon fusion can be
extensively enhanced by squark-pair threshold effects in the box-type 
diagrams, so that it can 
reach the size of 
the hadronic cross section via the subprocess
$b \overline{b} \to W^\pm H^\mp$ which appears at tree level.
\end{abstract}

\end{titlepage}


\section{Introduction}

In the Standard Model (SM), a single neutral Higgs boson is predicted
as a direct consequence of the SM mechanism of electroweak
symmetry breaking. The detection of this particle is, therefore,
one of the most essential tasks at
present and future collider experiments, especially at  
the Large Hadron Collider (LHC)~\cite{lhc}.
On the other hand, charged Higgs bosons (as well as a CP-odd neutral
Higgs boson) are predicted in extended versions of the SM model,
including the Minimal Supersymmetric Standard Model (MSSM).
Since a discovery of such an additional Higgs boson will 
immediately indicate physics beyond the SM, 
there is increasing interest in theoretical and experimental studies
to provide the basis for its accurate exploration.

At hadron colliders, a mode for the charged Higgs boson $H^\pm$
detection may be the top-antitop pair production from gluon fusion
and $q\bar{q}$ annihilation and subsequent decay
$t \to b H^+ \to b \tau^+ \nu$, 
if the mass of $H^\pm$ is smaller than $m_t - m_b$.
For heavier $H^\pm$, main modes for the $H^\pm$ production may be
those associated with heavy quarks, such as
$g b \to H^- t$~\cite{gbht}  
and $q b \to q' b H^-$~\cite{qbhqb}.
Although these processes give rather large production rates, 
they suffer from also large QCD backgrounds, especially when
the $H^\pm$ mass is above the threshold of $t\bar{b}$ pair production.
Pair production  of $H^\pm$ via the tree-level $q \overline{q}$
annihilation subprocesses 
and via the loop-induced gluon-fusion mechanism
have also been studied in the literature~\cite{bbhh,gghh,pphh,gghh-bh}.

A further possibility is single charged-Higgs-boson production
together with a charged $W$ gauge boson.
In this paper we focus on  the discussion of this process of
associate $W^\pm H^\mp$ production for the LHC. 
There are mainly two partonic 
subprocesses that contribute to the hadronic
cross section $pp \to W^-H^+$: 
the $b\bar{b}$ annihilation (at the tree level)
and the $gg$ fusion (at the one-loop level).
The production cross section based on
$gg \to W^\pm H^\mp$ can be comparable to that via
$b \overline{b} \to W^\pm H^\mp$, because
of the large number of gluons in the high energy proton
beams at the LHC.
The associate $W^\pm H^\mp$ production
via $b\overline{b}$ annihilation and
the $gg$ fusion  have been discussed at first in~\cite{pphw1}
for a supersymmetric 2-Higgs-doublet model including the loop contributions
from top and bottom quarks with the approximation $m_b = 0$. 
That work has been extended in~\cite{pphw-kniehl1}   
including a non-zero $b$-quark mass, thus allowing the investigation  
of the process for arbitrary values of $\tan\beta$.
The rate of $W^\pm H^\mp$ production mediated by the quark loops 
turned out to  be sizable, especially for low and high values 
of $\tan\beta$. 
The background, which mainly comes from $t\overline{t}$ production,  
has been analysed in \cite{mo}.  
These studies correspond to a MSSM scenario where the scalar quarks
are sufficiently heavy to decouple from the loop contributions.

In more general scenarios the squarks need not be heavy, and 
taking the MSSM seriously requires the inclusion of 
also the squark loops in the gluon-fusion mechanism 
and to study their effects on the predictions for $pp \to W^-H^+$. 
There are several reasons to underline the importance of the
squark contributions.
In the subprocess $gg \to W^\pm H^\mp$, the quark-loop
contributions are destructive between the triangle-type diagrams and
the box graphs~\cite{pphw1,pphw-kniehl1} in the MSSM; 
the effects from both types are almost of 
the same size, so that the cross section obtained from 
the summed $t$-$b$ loop contributions is more
than one order of magnitude smaller than that with only triangle-
or box-type diagrams separately. By this mechanism of cancellation 
in the quark-loop terms the relative importance  
of the squark-loop contribution increases. 
On top, the cross section for
$gg \to W^\pm H^\mp$ can be sizeably
enhanced by threshold effects of squark pairs, such that in special cases
the hadronic cross section via gluon fusion can be as large as
that via $b \overline{b} \to W^\pm H^\mp$.
Moreover, the present calculation of the
$b \overline{b}$ annihilation is expected to overestimate 
the cross section, as it has been mentioned \cite{pphw1,pphw-kniehl1}
by recalling the problem of double counting\cite{dble-count}.
Owing to the presence of also the superpartners in the virtual
states of the loop diagrams, the gluon-fusion mechanism is more
sensitive to the detailed structure of the model than the tree-level
process of $b\bar{b}$ annihilation.\footnote{ 
For the subprocess $b\overline{b}\to W^-H^+$, the electroweak one-loop 
corrections have recently been studied in Ref.~\cite{bbhw-ew}, which can 
give rise to a 10-15\% reduction of the lowest-order result.}

In this article we extend the previous calculations by including also
the scalar-quark sector in the loop diagrams for $gg$ fusion. 
We give analytical results and a detailed discussion of the effects 
in the hadronic cross section. Our results are for general parameters
of the MSSM;
for the numerical discussion, constraints from
the direct search  and from the precision data~\cite{pdg2000,hagiwara}
are taken into account.\footnote{
Quite recently, the squark-loop contributions to gluon fusion have
also been derived~\cite{pphw-kniehl2} with a study of their
effects in a supergravity-inspired GUT scenario.}

The paper is organized as follows.
In Section 2 the calculation for the partonic cross section, 
$gg \to W^-H^+$, is explained, and  
numerical results for the partonic and hadronic cross sections 
are presented in Section 3. Our conclusion is given 
in Section 4. 
All the relevant coupling constants and the analytic formulae for 
the amplitudes are presented in the Appendix for a 
comprehensive documentation.

\section{The partonic process $gg\to W^- H^+$}

\subsection{Cross section}\label{partcs}

In our kinematical conventions, the momenta of the initial state
gluons, $k$ and $\bar{k}$, are chosen as incoming and outgoing
for the momenta, $p$ and $\bar{p}$, of the final state 
particles:
\[
 g(k,a,\sigma) + g(\overline{k},b,\overline{\sigma})  
  \rightarrow W^-(p,\lambda) + H^+(\overline{p}) \, .
\]
Besides by their momenta, the initial state gluons are 
characterized by their color indices $a,b$ and their helicities
$\sigma,\bar{\sigma}$ ($= \pm 1$),  
and the final state $W$ boson is characterized 
by its helicity $\lambda$ ($= 0, \pm 1$).
We make use of the parton kinematical invariants
\begin{align*}
\hat{s}  = (k+\overline{k})^2\, , \quad 
\hat{t}  = (k-p)^2\, , \quad
\hat{u}  = (k-\overline{p})^2
\end{align*}
obeying the relation
\begin{align*}
\hat{s} + \hat{t} + \hat{u} = \mhpm^2 + m_W^2
\ehsp.
\end{align*}
The spin- and color-averaged cross section for the parton process
\begin{align}
\frac{d \sigma}{ d \hat{t}} & = \frac{1}{16\pi \hat{s}^2} \:
\sum_{\lambda = 0,\pm}^{}
\frac{1}{4} \sum_{\sigma,\bar{\sigma} = \pm 1}^{}
\frac{(\text{CF})}{64}
\big| \MM_{\sigma\bar{\sigma}\lambda} \big|^2, 
\;\; {\rm with}\;\; 
(\text{CF}) = \sum_{a,b = 1}^{8} 
\bigg[
\text{Tr} \big\{ \frac{\lambda^a}{2}
\frac{\lambda^b}{2} \big\}
\bigg]^2 = 2,
\end{align}
contains the helicity amplitudes
\begin{align}
\MM_{\sigma\bar{\sigma}\lambda} & =
\polv{\sigma}{\mu}{k} \polv{\bar{\sigma}}{\nu}{\overline{k}} 
\polv{\lambda}{\ast \rho}{p} \,
\widetilde{\MM}_{\mu\nu\rho} 
\ehsp, 
\end{align}
where $\polv{\sigma}{\mu}{k}$, 
      $\polv{\bar{\sigma}}{\nu}{\overline{k}}$ and  
      $\polv{\lambda}{\ast \rho}{p}$ 
are the polarization vectors for incoming gluons and 
outgoing $W$ bosons. 
As a general feature of the amplitude, 
the transversality of gluons gives useful identities 
\begin{align}
k^\mu \polv{\bar{\sigma}}{\nu}{\overline{k}} 
\polv{\lambda}{\ast \rho}{p} \,
\widetilde{\MM}_{\mu\nu\rho} 
=
\polv{\sigma}{\mu}{k} \overline{k}^\nu 
\polv{\lambda}{\ast \rho}{p} \,
\widetilde{\MM}_{\mu\nu\rho} = 0
\ehsp,  \label{inv}
\end{align}
which allow cross-checks of our one-loop calculation of 
$\widetilde{\MM}_{\mu\nu\rho}$.

In the parton Center-of-Mass (CM) frame, the momenta may be expressed by 
\begin{align}
k^\mu&=
\left(\frac{\sqrt{\hat{s}}}{2},0,0,\frac{\sqrt{\hat{s}}}{2}\right), \\
\overline{k}^\mu&=
\left(\frac{\sqrt{\hat{s}}}{2},0,0,-\frac{\sqrt{\hat{s}}}{2}\right), \\
p^\mu&= (E_W, \vec{p}_W) =
\left(E_W,|\vec{p}_W|\sin\Theta,0,|\vec{p}_W|\cos\Theta\right),
\end{align}
and then the polarization vectors are given by  
\begin{align}
\polv{\sigma}{\mu}{k}&=
\frac{1}{\sqrt{2}} 
\left(0,1 ,i \sigma , 0 \right), \\
\polv{\overline{\sigma}}{\mu}{\overline{k}}&=
\frac{1}{\sqrt{2}}
\left(0,1 ,-i \overline{\sigma} , 0 \right), 
\end{align}
and
\begin{align}
 \polv{\lambda=0}{\ast \mu}{p}&=
\left(\frac{|\vec{p}_W|}{m_W},\frac{E_W}{m_W}\sin\Theta,0,
                                \frac{E_W}{m_W}\cos\Theta\right), \\ 
 \polv{\lambda=\pm}{\ast \mu}{p}&=
\frac{1}{\sqrt{2}}
\left(0,i \lambda\cos\Theta,1,-i\lambda\sin\Theta \right).  
\end{align}

Finally, the integrated partonic cross section
\begin{align}
\sigma_{gg \to W^- H^+}(\hat{s},\alpha_S(\mu_R)) & =
\int_{ \hat{t}_{\text{min}}(\hat{s}) }^{ \hat{t}_{\text{max}}(\hat{s}) } 
\!\!\! d\hat{t}
\: \frac{d \sigma}{ d \hat{t}}
\end{align}
is evaluated by numerical integration over the 
kinematically allowed $\hat{t}$-range 
for a given CM-energy $\sqrts$.
The renormalization scale $\mu_R$ in the $\overline{\rm MS}$ 
strong coupling constant $\alpha_S(\mu)$ 
is chosen as $\sqrts$, the hard energy scale of the process.

\subsection{Calculation of the amplitude}

In the MSSM, there is no tree-level contribution to the subprocess 
$gg \to W^-H^+$. 
The process is induced at the one-loop level by diagrams with  
quark-loops (see Appendix~\ref{q-graphs}) and with squark-loops  
(see Appendix~\ref{sq-graphs}). 
In the loop diagrams, also the neutral Higgs bosons $h^0$, $H^0$ and $A^0$ 
of the MSSM appear as well as the scalar partners of the quarks. 
Notations and couplings are collected in Appendix \ref{couplings}. 

The quark-loop diagrams can be subdivided into box-type diagrams and into
triangle diagrams with s-channel exchange of a neutral Higgs boson. 
In each box diagram, the quarks in the loop couple directly to the outgoing
charged Higgs boson $H^+$, while in the triangle diagrams the 
quarks couple to one of the neutral Higgs bosons $h^0$, $H^0$ and $A^0$. 
Since these Yukawa interactions are all proportional to the quark 
masses, the contributions from the loop diagrams of the 
third-generation quarks $(t,b)$ are dominant.

It has been known that there is a striking feature among the 
quark-loop contributions to the amplitude in the MSSM. 
Owing to supersymmetry, the box-diagram contribution to the amplitude and 
the triangle-type one are almost of the same size but the relative sign is 
negative, so that strong destructive interference occurs. 
The cross section resulting from all the quark-loop diagrams 
is therefore one to two orders of magnitude smaller than the cross section 
obtained from only box-type diagrams or only triangle diagrams, separately.

By the destructive interference in the quark-loop contributions,  
the squark-loop effects become relatively large. 
The squark one-loop diagrams are subdivided into 
(1) diagrams with a 2-point loop and an intermediate 
s-channel neutral CP-even 
 Higgs boson,   
(2) diagrams with a triangle and an intermediate neutral CP-even Higgs boson,   
(3) diagrams with a triangle without an intermediate Higgs boson,   
and  
(4) box-type diagrams.
As has been verified explicitly, each group separately
fulfills the transversality relations~(\ref{inv}).

The cross section has been derived in two completely independent
calculations and perfect agreement has been achieved.
Furthermore, these results have been confirmed by utilizing
FeynArts and FormCalc \cite{FAFC}.
The agreement of these three independent calculations and the
transversality test establish strong confidence in our result.

\section{Numerical Results}

\subsection{Parameters}\label{parameters}

We take $m_Z^{}$, $m_W^{}$ and $G_F^{}$ 
as the input electroweak parameters, and use values 
$m_Z=91.1882$ GeV, $m_W=80.419$ GeV and $G_F=1.16639 \times 10^{-5}$ 
GeV$^{-2}$~\cite{pdg2000}. 
For the strong coupling constant $\alpha_S(\mu_R)$, we use the formula 
including the two-loop QCD corrections for $n_f=5$ with 
$\Lambda^5_{QCD}=170$ MeV which can be found in~\cite{pdg2000}.
The mass of the top and bottom quarks are fixed here as $m_t=174.3$ GeV 
and $m_b=4.7$ GeV. 
  
The sfermion parameter sets are chosen in accordance with  
the direct search results~\cite{pdg2000}. 
In addition, 
the stringent experimental constraints on the new physics parameters 
from the electroweak precision measurements are respected. 
In Refs.~\cite{hagiwara,eeww}, 
this kind of constraint on the new physics parameters 
has been studied in the framework of the MSSM by using the $Z$-pole 
data, the $m_W$ measurements, and low-energy neutral-current data. 
Our parameter sets are chosen to be in accordance with the region 
inside the 99\% CL contour on the $S_Z$-$T_Z$ plane 
presented in~\cite{eeww}. 

In cases without sfermion mixing, 
it is expected that lighter sfermions give larger 
one-loop contributions while heavy sfermions tend to 
decouple from the observables.    
We examine the cross section under this situation 
by introducing three cases, specified as  A, B, C in Table 1. 
Another interesting situation is the case with large 
$\tilde{t}_L$-$\tilde{t}_R$ mixing. 
The magnitude of the mixing is determined by the off-diagonal part 
of the $\tilde{t}$-mass matrix, especially $X_t$ (see Appendix A). 
To study the cross section in this situation, we select 
three parameter sets (Case 1, 2, 3 in Table 2), in which the maximal 
$\tilde{t}_L$-$\tilde{t}_R$ mixing occurs with the mixing angle 
$\theta_{\tilde{t}} \sim 45^\circ$ and a light $\tilde{t}_1$ with 
$m_{\tilde{t}_1} \sim 100$ GeV.

\begin{table}[hbt]
\begin{center}
\begin{tabular}[10]{|l||c|c|c||c|c|c||c|c|c|}
\hline
&\multicolumn{3}{|c||}{\bf Case A} &\multicolumn{3}{|c||}{\bf Case B} &\multicolumn{3}{|c|}{\bf Case C}\\
\hline
$\MSQ \: [\gev]$ & \multicolumn{3}{|c||}{250} &\multicolumn{3}{|c||}{300} &\multicolumn{3}{|c|}{350}\\
\hline
$\MSU \: [\gev]$ & \multicolumn{3}{|c||}{250} &\multicolumn{3}{|c||}{300} &\multicolumn{3}{|c|}{350}\\
\hline
$\MSD \: [\gev]$ & \multicolumn{3}{|c||}{250} &\multicolumn{3}{|c||}{300} &\multicolumn{3}{|c|}{350}\\
\hline
$X_t \: [\gev]$  & \multicolumn{3}{|c||}{0}   &\multicolumn{3}{|c||}{0}   &\multicolumn{3}{|c|}{0}\\
\hline
$X_b \: [\gev]$  & \multicolumn{3}{|c||}{0}   &\multicolumn{3}{|c||}{0}   &\multicolumn{3}{|c|}{0}\\
\hline
$\tan\beta$      & 1.5     & 6       & 30     & 1.5     & 6      & 30     & 1.5     & 6      & 30\\
\hline\hline 
$m_{\Stop_1} \: [\gev]$ 
                 & 303     & 300     & 300    & 345     & 343    & 343    & 390     & 387    & 387\\
\hline
$m_{\Stop_2} \: [\gev]$ 
                 & 304     & 303     & 303    & 346     & 345    & 345    & 390     & 390    & 389\\
\hline 
$m_{\Sbot_1} \: [\gev]$ 
                 & 250     & 251     & 251    & 300     & 301    & 301    & 350     & 351    & 351\\
\hline
$m_{\Sbot_2} \: [\gev]$ 
                 & 253     & 257     & 257    & 302     & 306    & 306    & 352     & 355    & 355\\
\hline
\end{tabular}
\caption{\label{nomixing-cases}
        Choices for the the soft supersymmetry breaking parameters $\MSQ,\MSU,
        \MSD,X_t$ and $X_b$ without squark mixing (i.e. $X_t = X_b = 0$).
        For all cases, $\mu$ is fixed to be zero.   
        The resulting spectrum of the third generation squark masses is 
        also displayed.  
        }
\end{center}
\end{table}

\begin{table}[hbt]
\begin{center}
\begin{tabular}[10]{|l||c|c|c||c|c|c||c|c|c|}
\hline
&\multicolumn{3}{|c||}{\bf Case 1} &\multicolumn{3}{|c||}{\bf Case 2} &\multicolumn{3}{|c|}{\bf Case 3}\\
\hline
$\MSQ \: [\gev]$ & \multicolumn{3}{|c||}{250} &\multicolumn{3}{|c||}{300} &\multicolumn{3}{|c|}{350}\\
\hline
$\MSU \: [\gev]$ & \multicolumn{3}{|c||}{250} &\multicolumn{3}{|c||}{300} &\multicolumn{3}{|c|}{350}\\
\hline
$\MSD \: [\gev]$ & \multicolumn{3}{|c||}{250} &\multicolumn{3}{|c||}{300} &\multicolumn{3}{|c|}{350}\\
\hline
$X_b \: [\gev]$  & \multicolumn{3}{|c||}{0}   &\multicolumn{3}{|c||}{0}   &\multicolumn{3}{|c|}{0}\\
\hline
$\tan\beta$      & 1.5    & 6      & 30      & 1.5    & 6      & 30     & 1.5    & 6      & 30\\
\hline
$X_t \: [\gev]$  & -470   & -464   & -463    & -628   & -621   & -620   & -813   & -806   & -806\\
\hline\hline
$m_{\Stop_1} \: [\gev]$ 
                 & 101    & 100    & 101     & 101    & 101    & 101    & 102    & 102    & 102\\
\hline
$m_{\Stop_2} \: [\gev]$ 
                 & 417    & 414    & 414     & 479    & 476    & 476    & 542    & 540    & 540\\
\hline
$m_{\Sbot_1} \: [\gev]$ 
                 & 251    & 251    & 251     & 300    & 301    & 301    & 350    & 351    & 351\\
\hline
$m_{\Sbot_2} \: [\gev]$
                 & 253    & 257    & 257     & 302    & 306    & 306    & 352    & 355    & 355\\
\hline
$\theta_{\Stop} \: [^\circ]$ 
                 & 44.9   & 44.7   & 44.7    & 44.9   & 44.8   & 44.8   & 44.9   & 44.8   & 44.8\\ 
\hline
\end{tabular}
\caption{\label{mixing-cases}
        Choices for the the soft supersymmetry breaking parameters $\MSQ,\MSU,
        \MSD,X_t$ and $X_b$ for three values of $\tan\beta$ and the 
        resulting spectrum of the third-generation squark masses and
        the mixing angle in the $\Stop$-sector. 
        For all cases, $\mu$ is fixed to be zero.   
        }
\end{center}
\end{table}

\subsection{Partonic Cross Section}

Before proceeding to the hadronic cross section, $pp \to W^-H^+$, 
we want to illustrate the partonic cross section of the subprocess 
$gg \to W^-H^+$ numerically. 
Although the partonic process is not accessible experimentally, 
it is useful for understanding the features of the hadronic 
cross section, in particular its dependence on the MSSM parameters. 
For gluon-fusion processes, the threshold region    
($\sqrt{\hat{s}} \sim m_W^{}+m_{H^\pm}$), 
where gluon pairs are most numerous in proton collisions, 
gives the dominant contribution to the hadronic cross section.
Not intending completeness, we display here only one 
parameter set (Case 1 of Table 2) for the squark sector as an example. 

\begin{figure}[bh]
\begin{center}
  \psfrag{SQRTS}[c]{$\sqrts \; [\gev]$}
  \psfrag{SIGMA}[c]{$\sigma( gg \to H^+ W^- ) \; [\fb]$}
  \psfrag{MHPM100}{\scriptsize $ \mhpm = 100\,\gev$}
  \psfrag{MHPM210}{\scriptsize $ \mhpm = 210\,\gev$}
  \psfrag{MHPM360}{\scriptsize $ \mhpm = 360\,\gev$}
  \psfrag{MHPM450}{\scriptsize $ \mhpm = 450\,\gev$}
  \psfrag{FIGA}{\scriptsize (a)}
  \psfrag{FIGB}{\scriptsize (b)}
  \psfrag{FIGC}{\scriptsize (c)}
  \psfrag{FIGD}{\scriptsize (d)} 
\resizebox*{1.2\width}{1.2\height}{\includegraphics*{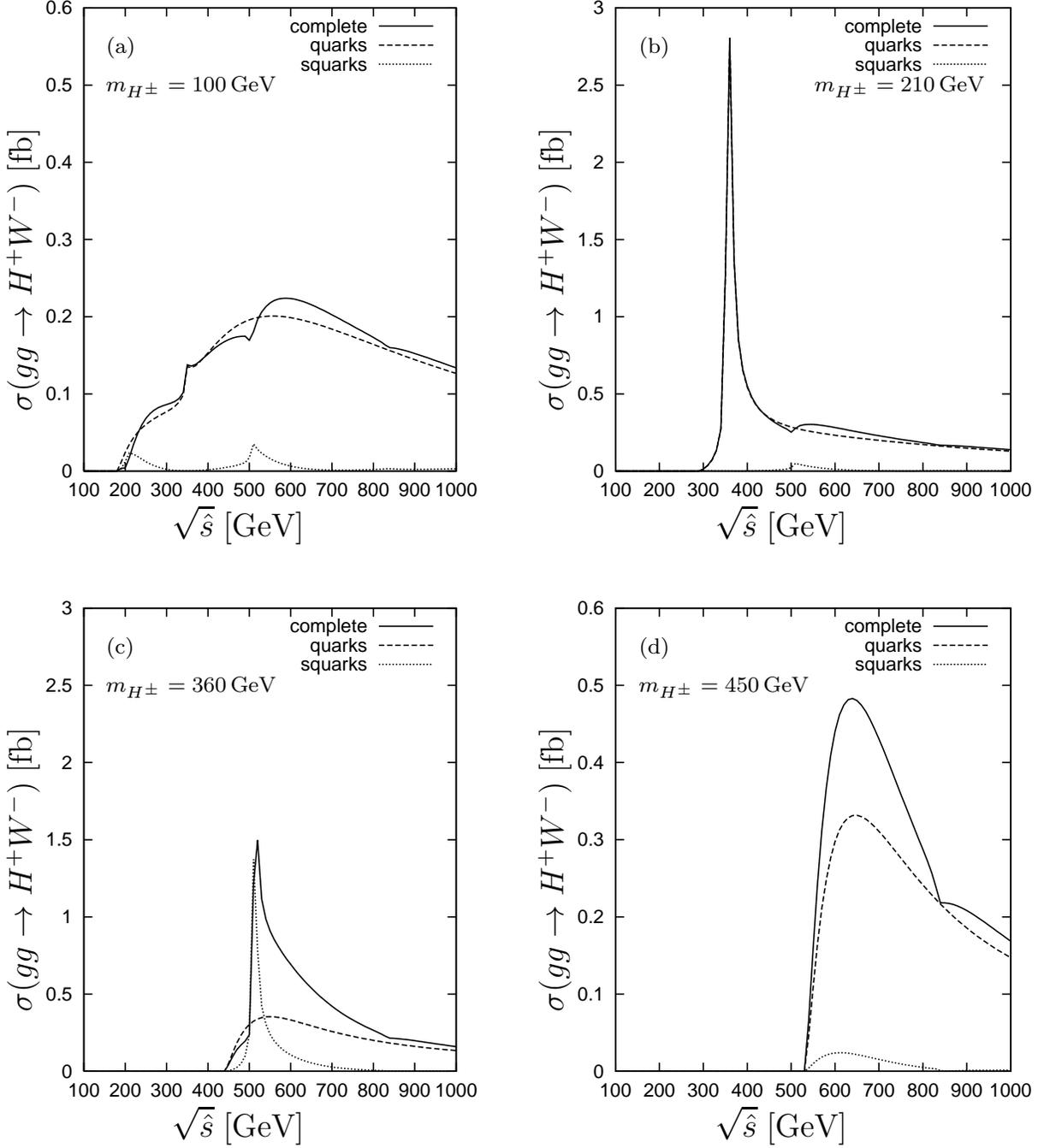}}
    \caption{\label{parton-cs} Partonic cross section evaluated using
        all Feynman graphs (solid lines), only quark loop graphs 
        (dashed lines) and only squark loop graphs (dotted lines) 
        for a sequence of charged-Higgs masses 
        ($100,210,360,450 \,\gev$) and $\tan\beta = 1.5$.
        The parameter set 1 (see table \ref{mixing-cases}) 
        is used in the squark sector.
            }
  \end{center}
\end{figure}

In Figure \ref{parton-cs} the integrated partonic cross section
is shown as a function of $\sqrts$ for a sequence of $\mhpm$ values. 
Each plot displays three curves: the cross section evaluated from all 
(solid lines), only squark- (dotted lines) and 
only quark-loop diagrams (dashed lines). 
The last case corresponds 
to the scenario where the squarks decouple.

For $\mhpm = 100 \,\gev$ [Figure 1(a)] 
the cross section is mainly dominated
by the quark-loop diagrams.
The $\tilde{b}\tilde{b}$ 
thresholds at $\sqrts \sim 2 m_{\Sbot_1} \sim 2 m_{\Sbot_2}$ 
from the squark-box diagrams are visible.
Nevertheless, the
$\tilde{b}\tilde{b}$ thresholds turn out to be 
of less importance for the hadronic cross section, because they are 
too far off the threshold of the associated $H^+ W^-$ production. 
Another effect of the  
squark-loop diagrams is present
near the production threshold, where the tail of the 
$\tilde{t}_1\tilde{t}_1$-threshold still has influence and gives
rise to a slightly enhanced hadronic cross section with respect to
the squark decoupling case (see Figure \ref{mhpm-mixing}).

At a charged-Higgs-boson mass of $210 \,\gev$ [Figure 1(b)] 
the cross section is largely 
dominated by the quark-loop diagrams. The peak corresponds to  
the $t\overline{t}$-threshold effect in the quark-box diagrams.
A constructive squark-loop effect appears above the sbottom pair 
threshold, but this is negligible in the hadronic cross section.

For $\mhpm = 360 \,\gev$ [Figure 1(c)]  
a novel enhancement of the partonic cross section 
occurs around the thresholds of sbottom pair production 
$\sqrts \sim 2 m_{\Sbot_1} \sim 2 m_{\Sbot_2}$, 
where the squark-loop contributions are 
dominant. Since the peak in the partonic cross section is near the $W^-H^+$ 
threshold where a lot of gluons are supplied, a large squark contribution 
appears also in the hadronic cross section, which corresponds to the local 
maximum of the hadronic cross section at around $m_{H^\pm}=360$ GeV 
in Figure \ref{mhpm-mixing}.

For $\mhpm = 450 \,\gev$ [Figure 1(d)]  
there is neither a quark-pair nor a squark-pair threshold 
slightly above the $W^-H^+$ threshold, but the enhancement 
due to virtual squarks is still quite relevant.

\subsection{Hadronic Cross Section} 

The hadronic inclusive cross section for $W^- H^+$ production in 
proton-proton collisions at a total hadronic CM energy 
$\sqrt{S}$ can be written as a convolution \cite{pQCD-lect}
\begin{align}
\label{hadronx}
\sigma( p p  \to W^- H^+ + X) & =
\sum_{\{n,m\}}^{} 
\int_{\tau_0}^{1} d\tau \frac{ d{\cal L}^{pp}_{nm} }{ d\tau }
\;\sigma_{n m \to W^- H^+}(\tau S,\alpha_S(\mu_R))
\end{align}
with the parton luminosity
\begin{align}
\frac{ d{\cal L}^{pp}_{nm} }{ d\tau } & =
        \int_{\tau}^{1} \frac{dx}{x}
        \frac{1}{1+\delta_{nm}} 
        \Big[
        f_{n/p} (x,\mu_F) f_{m/p} (\frac{\tau}{x},\mu_F)
        + f_{m/p} (x,\mu_F) f_{n/p} (\frac{\tau}{x},\mu_F)
        \Big]
        \, ,
\end{align}
where $f_{n/p} (x,\mu_F)$ denotes the density of partons of type $n$
in the proton carrying a fraction $x$ of the proton momentum
at the scale $\mu_F$. 
In our case the sum over unordered pairs of partons $\{n,m\}$ 
reduces to two terms, i.e. there are two parton subprocesses
contributing to inclusive $W^- H^+$ hadroproduction; gluon fusion
and $b \bar{b}$-annihilation. 
Our main concern is the gluon-fusion
process and the possible enhancement of the cross section
through virtual squark effects
with respect to the approximation of decoupling squarks.
We will also compare the cross section for gluon fusion
with the one for $b \bar{b}$-annihilation.
The numerical evaluation has been carried out with the MRS(G)
gluon distribution functions \cite{MRSG} and with the renormalization
and factorization scale $\mu_R,\mu_F$ chosen as equal.


The input parameters in the MSSM Higgs sector can be chosen to
be $\mhpm$ and $\tan\beta$. Thus we study
here the dependence of the hadronic cross section for $W^- H^+$
production via gluon fusion on these two parameters for 
the different squark scenarios mentioned in chapter 
\ref{parameters}. 

\subsubsection{Unmixed sfermions}
%

\begin{figure}[hbt]
\begin{center}
  \psfrag{SQRTS}[c]{$\sqrts \; [\gev]$}
  \psfrag{SIGMAPP}[c]{$\sigma( pp \to H^+ W^- ) \; [\fb]$}
  \psfrag{TANB1.5}{$ \tan\beta = 1.5$}
  \psfrag{TANB006}{$ \tan\beta = 6$}
  \psfrag{TANB030}{$ \tan\beta = 30$}
  \psfrag{MHPM}[c]{$\mhpm \; [\gev]$}
  \psfrag{BBBAR}[lb]{via $b \bar{b}$ }
  \resizebox*{1.5\width}{1.5\height}{\includegraphics*{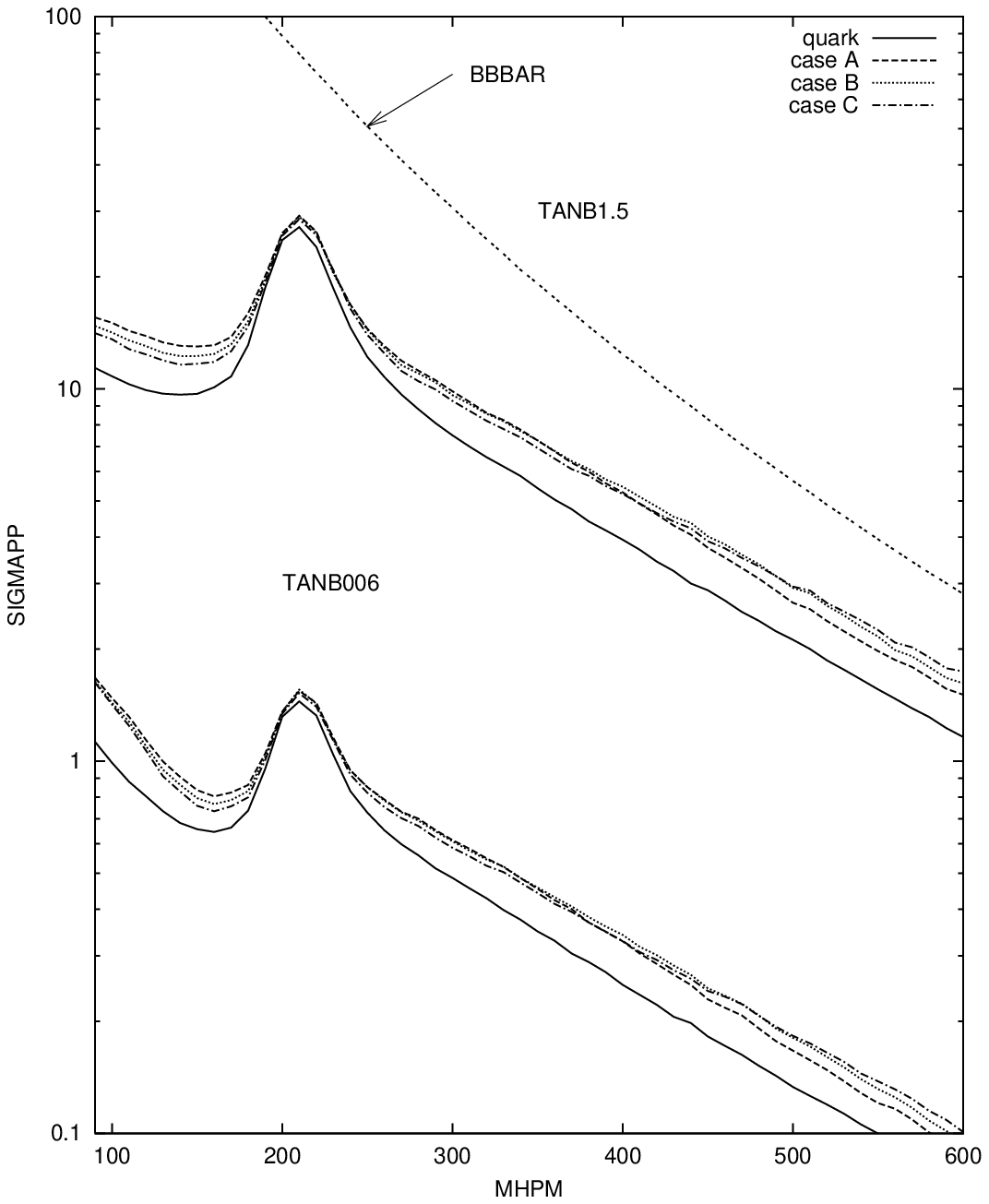}}
    \caption{\label{mhpm-nomixing} Hadronic cross section for $H^+ W^-$
        production via gluon fusion versus the charged-Higgs
        mass $\mhpm$ for two values of $\tan\beta$ (1.5,6).
        The three squark scenarios without mixing (A, B, C in Table 
        \ref{nomixing-cases}) are compared with the case of decoupling
        squarks (solid lines).
            }
  \end{center}
\end{figure}

Figure \ref{mhpm-nomixing} shows the variation of 
the hadronic cross section\footnote{With the 
'hadronic cross section' we denote 
henceforth the cross section for $W^- H^+$
production via gluon fusion and take care that no
confusion arises when the hadronic cross section for production
of $W^- H^+$ via $b \bar{b}$-annihilation is addressed.}
with $\mhpm$ for the squark scenarios without mixing (cases A, B, C, 
see Table \ref{nomixing-cases}) and for two values of 
$\tan\beta$ (1.5 and 6). The cross section is decreasing
rapidly with increasing charged-Higgs mass, except for the region
around $\mhpm = 210~\gev$ where a peak appears. The decrease of 
the cross section comes from the gluon luminosity, while the 
rise of the cross section in the peak area is due to the top-pair 
threshold in the quark-box diagrams, which gives rise 
to a sharp peak in the partonic cross section 
(see [Figure \ref{parton-cs}(b)]) at $\sqrts = 2 m_t$ 
when the production threshold for the $W^-H^+$ 
is near $2 m_t$.
The squark scenarios without mixing (cases A, B, C, see Table 1) 
show generally an enhancement 
of the cross section with respect to the decoupling case 
of about 25\% to 35\% over the depicted 
$\mhpm$-range, except for the peak region between about $180\,\gev$
and $250\,\gev$. Clearly, in this range the quark loop graphs dominate
and the squark contribution corrects the quark-loop result only by a
few percent. 

\begin{figure}[hbt]
\begin{center}
  \psfrag{SIGMAPP}[c]{$\sigma( pp \to H^+ W^- ) \; [\fb]$}
  \psfrag{TANB}{$\tan\beta$}
  \psfrag{MHPM100}[l]{$\mhpm=100\,\gev$}
  \psfrag{MHPM300}[l]{$\mhpm=300\,\gev$}
  \psfrag{MHPM1000}[l]{$\mhpm=1000\,\gev$}
\resizebox*{1.5\width}{1.5\height}{\includegraphics*{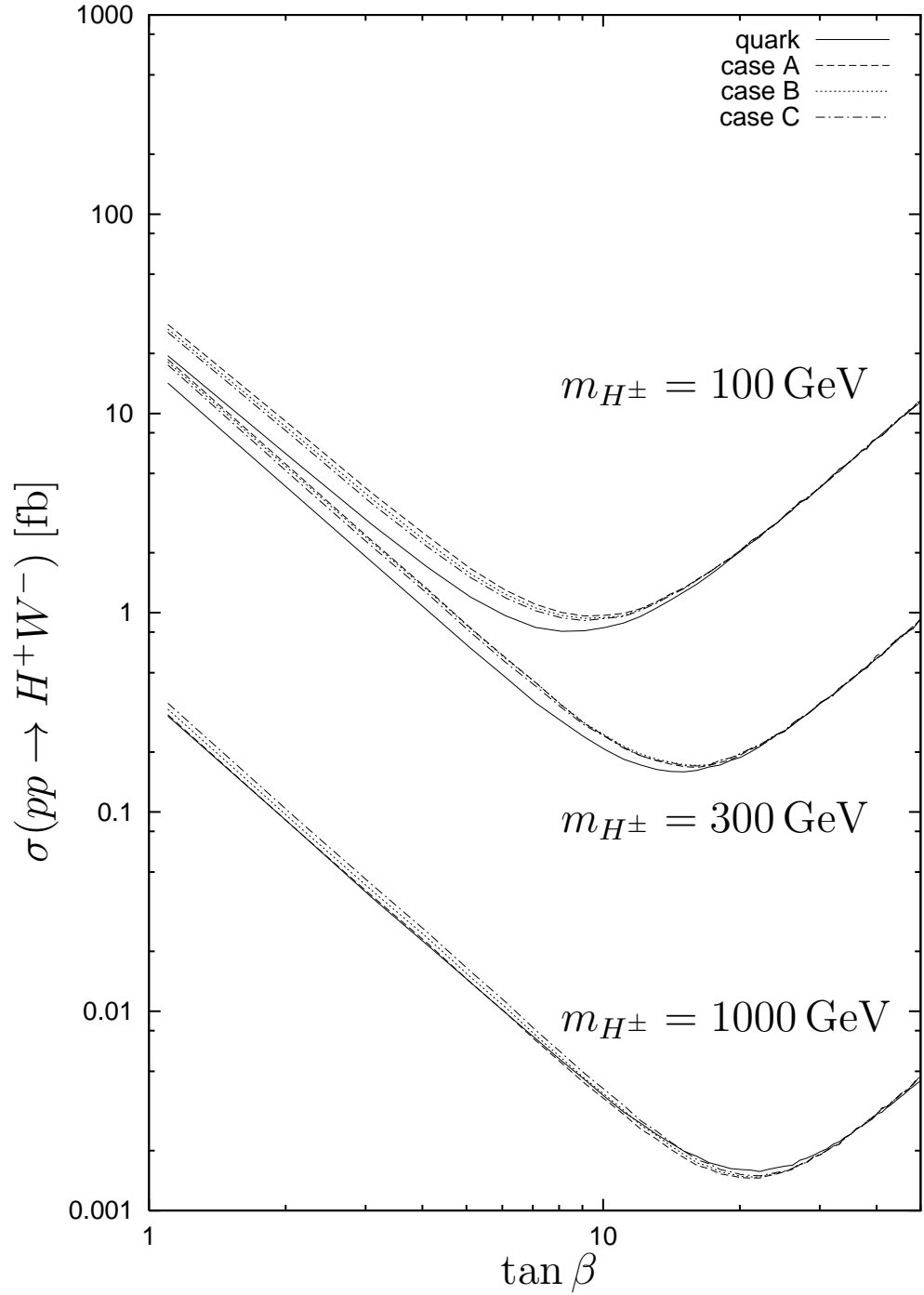}}
    \caption{\label{tb-nomixing} Hadronic cross section for $H^+ W^-$
        production via gluon fusion versus $\tan\beta$ for 
        three values of the charged-Higgs mass $\mhpm$ (100,300,1000 $\gev$).
        The three squark cases without mixing (A, B, C in Table
        \ref{nomixing-cases}) are compared 
        with the case of decoupling squarks (solid lines).
            }
  \end{center}
\end{figure}

The $\tan\beta$-dependence of the hadronic cross section 
is shown in Figure \ref{tb-nomixing} for the non-mixing 
cases (case A, B, C, Table \ref{nomixing-cases}). 
The case in which all squarks decouple is also shown for comparison.  
One can see that the squark-loop contributions enhance the 
hadronic cross section for $\tan\beta$ values mainly 
below the point where the cross section takes its minimum, 
The enhancement is bigger for smaller $\mhpm$, e.g. 40\% for squark-case A 
at $\mhpm=100\,\gev$ and $\tan\beta = 5$. 
For $\mhpm=100\,\gev$ and $300\,\gev$ the enhancement is 
highest for the squark-case A  
with the lowest squark mass scale, and lowest for the squark-case C
with the highest squark mass scale, while the situation is just
the opposite at $\mhpm=1000\,\gev$. 

\subsubsection{Maximal $\tilde{t}_L$-$\tilde{t}_R$ mixing}
%

\begin{figure}[hbt]
\begin{center}
  \psfrag{SQRTS}[c]{$\sqrts \; [\gev]$}
  \psfrag{SIGMAPP}[c]{$\sigma( pp \to H^+ W^- ) \; [\fb]$}
  \psfrag{TANB1.5}{$ \tan\beta = 1.5$}
  \psfrag{TANB006}{$ \tan\beta = 6$}
  \psfrag{TANB030}{$ \tan\beta = 30$}
  \psfrag{MHPM}[c]{$\mhpm \;[\gev]$}
  \psfrag{BBBAR}[lb]{via $b \bar{b}$ }
  \resizebox*{1.5\width}{1.5\height}{\includegraphics*{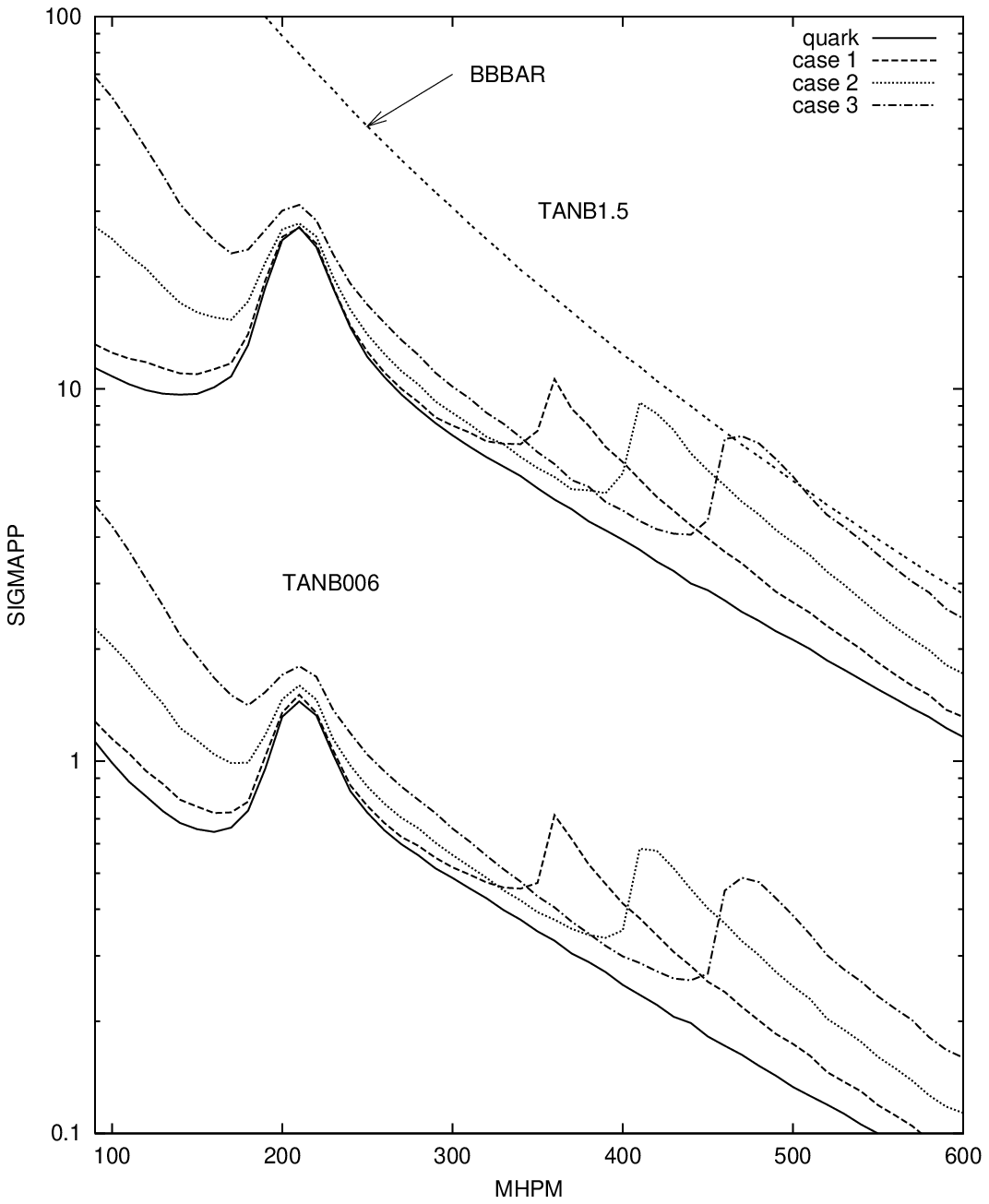}}
    \caption{\label{mhpm-mixing} Hadronic cross section for $H^+ W^-$ 
        production via gluon fusion versus the charged-Higgs
        mass $\mhpm$ for two values of $\tan\beta$ (1.5,6).
        The three squark scenarios with large mixing  (1, 2, 3 in Table 
        \ref{mixing-cases}) are compared with the case of decoupling
        squarks (solid lines).
            }
  \end{center}
\end{figure}

In Figure \ref{mhpm-mixing} 
the hadronic cross section for the squark scenarios with 
maximal $\tilde{t}_L$-$\tilde{t}_R$ mixing 
(cases 1, 2, 3, see Table \ref{mixing-cases}) 
is shown as a function of $m_{H^\pm}$ for $\tan\beta = 1.5$ and 6, 
comparing the three squark scenarios with the squark decoupling case. 
The general feature of all three scenarios is that there is 
a second peak besides the one due to quark-box diagrams. 
This second peak originates  
from the squark-pair threshold effects in the squark-loop diagrams. 
If the $W^- H^+$ production threshold is somewhat below 
$2 m_{\Sbot_{1(2)}}$, then the bottom-squark threshold of the 
box diagrams results in a pronounced peak in the parton 
cross section at $\sqrts = 2 m_{\Sbot_{1(2)}}$, which is even slightly higher
than the peak due to quark loops (see Figure \ref{parton-cs} (b) and (c)).
The magnitude of the peak can be traced back to strongly enhanced 
couplings of the charged Higgs to
$\Stop_1 \Sbot_{1}$ and $\Stop_2 \Sbot_{1}$ 
involving the non-diagonal entries of the $\tilde{t}$-mass matrix 
(see appendix \ref{couplings}).
It turns out that in the region of low $\tan\beta$ the contribution
to the hadronic cross section from gluon fusion can be comparable 
and even slightly larger than the
contribution from $b \bar{b}$ annihilation in the MSSM.
This important feature is also shown  
in Figure \ref{mhpm-mixing}, where the 
hadronic cross section for $W^- H^+$ production via 
$b \bar{b}$ annihilation for $\tan\beta = 1.5$ is also depicted 
for comparison.

\begin{figure}[hbt]
\begin{center}
  \psfrag{SIGMAPP}[c]{$\sigma( pp \to H^+ W^- ) \; [\fb]$}
  \psfrag{TANB}{$\tan\beta$}
  \psfrag{MHPM100}[l]{$\mhpm=100\,\gev$}
  \psfrag{MHPM360}[l]{$\mhpm=360\,\gev$}
  \psfrag{MHPM1000}[l]{$\mhpm=1000\,\gev$}
  \psfrag{BBBAR360}[r]{{\scriptsize via $b \bar{b}$ [$\mhpm=360\,\gev$]}}
\resizebox*{1.5\width}{1.5\height}{\includegraphics*{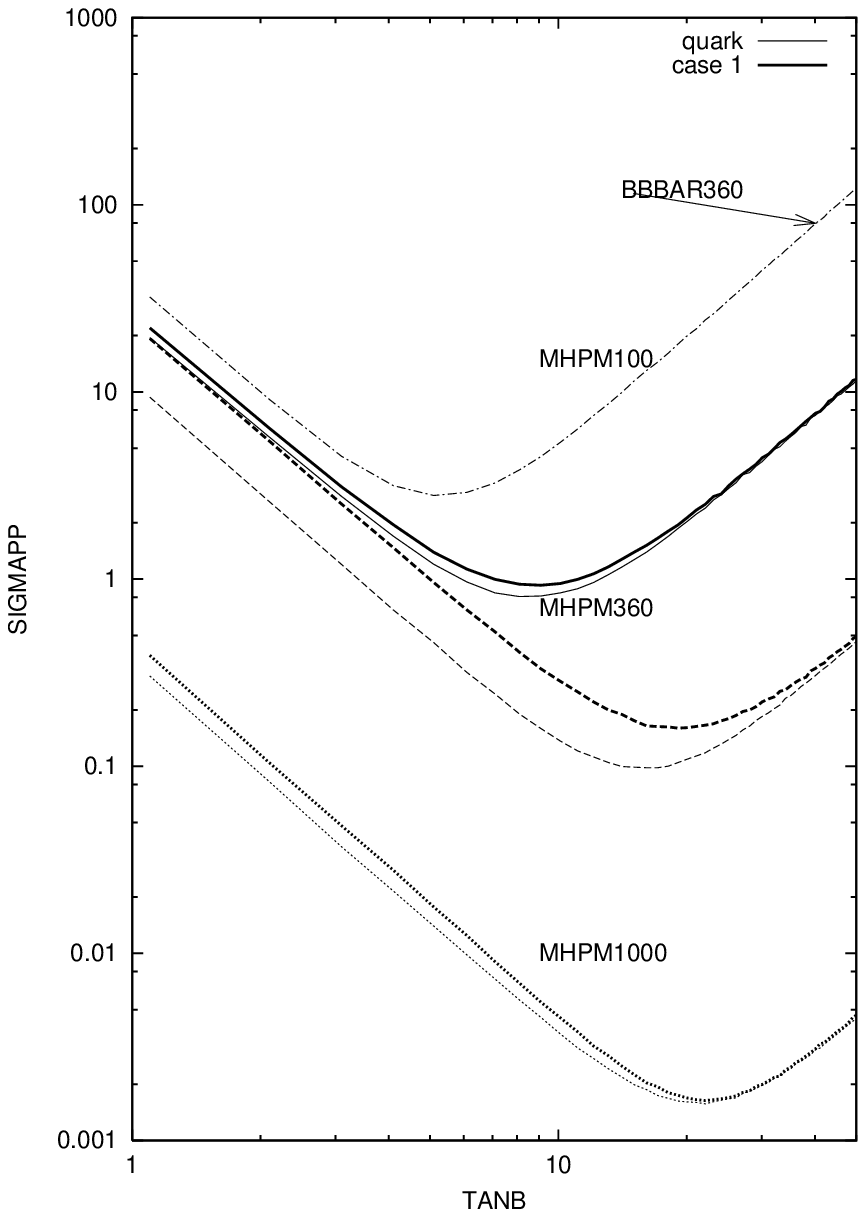}}
    \caption{\label{tb-case1} Hadronic cross section for $H^+ W^-$
        production via gluon fusion versus $\tan\beta$ for 
        three values of the charged-Higgs mass $\mhpm$ (100,360,1000 $\gev$). 
        The squark case 1 (thick lines) is compared to the case of 
        decoupling squarks (thin lines).
            }
  \end{center}
\end{figure}

\begin{figure}[hbt]
\begin{center}
  \psfrag{SIGMAPP}[c]{$\sigma( pp \to H^+ W^- ) \; [\fb]$}
  \psfrag{TANB}{$\tan\beta$}
  \psfrag{MHPM}[c]{$\mhpm \;\gev$}
  \psfrag{MHPM100}[l]{$\mhpm=100\,\gev$}
  \psfrag{MHPM410}[l]{$\mhpm=410\,\gev$}
  \psfrag{MHPM1000}[l]{$\mhpm=1000\,\gev$}
  \psfrag{BBBAR410}[r]{{\scriptsize via $b \bar{b}$ [$\mhpm=410\,\gev$]}}
\resizebox*{1.5\width}{1.5\height}{\includegraphics*{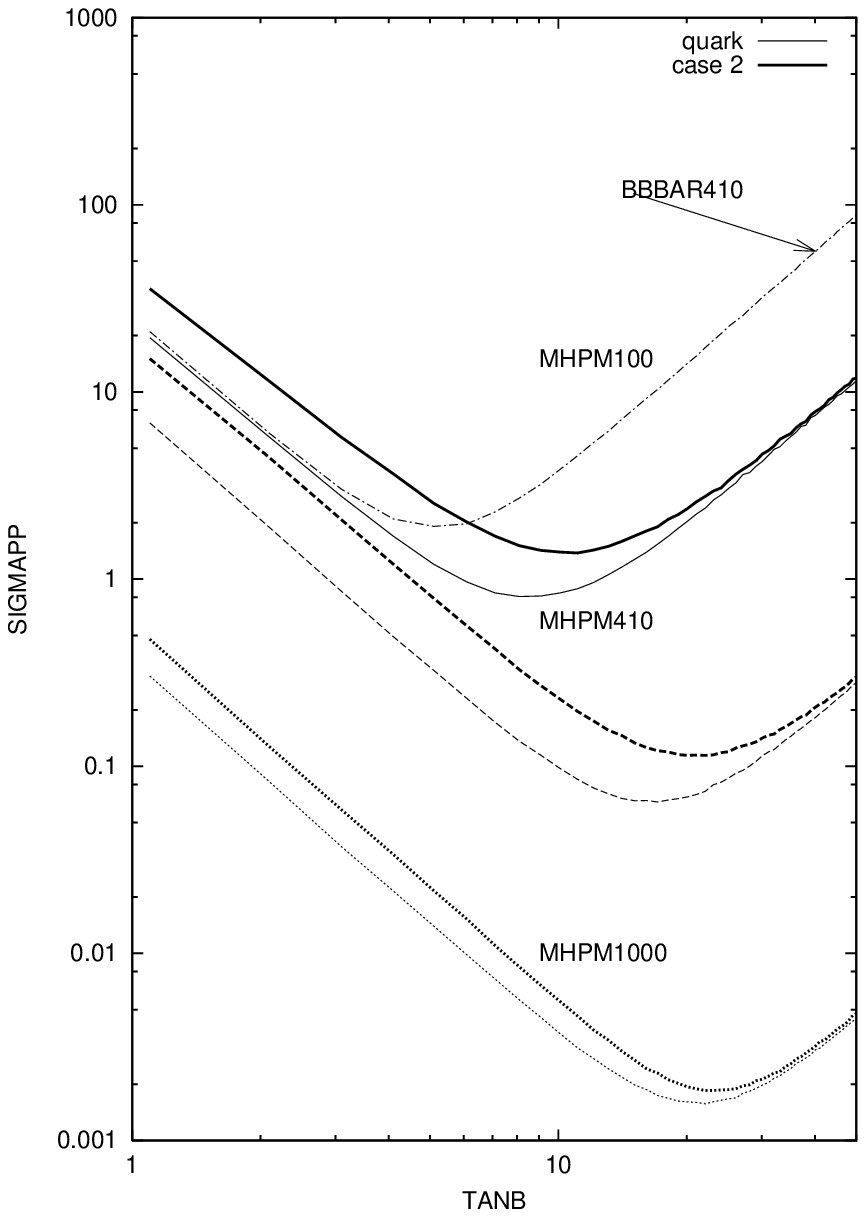}}
    \caption{\label{tb-case2} Hadronic cross section for $H^+ W^-$
        production via gluon fusion versus $\tan\beta$ for 
        three values of the charged-Higgs mass $\mhpm$ (100,410,1000 $\gev$). 
        The squark case 2 (thick lines) is compared to the case of 
        decoupling squarks (thin lines).
            }
  \end{center}
\end{figure}

\begin{figure}[hbt]
\begin{center}
  \psfrag{SIGMAPP}[c]{$\sigma( pp \to H^+ W^- ) \; [\fb]$}
  \psfrag{TANB}{$\tan\beta$}
  \psfrag{MHPM100}[l]{$\mhpm=100\,\gev$}
  \psfrag{MHPM470}[l]{$\mhpm=470\,\gev$}
  \psfrag{MHPM1000}[l]{$\mhpm=1000\,\gev$}
  \psfrag{BBBAR470}[r]{{\scriptsize via $b \bar{b}$ [$\mhpm=470\,\gev$]}}
\resizebox*{1.5\width}{1.5\height}{\includegraphics*{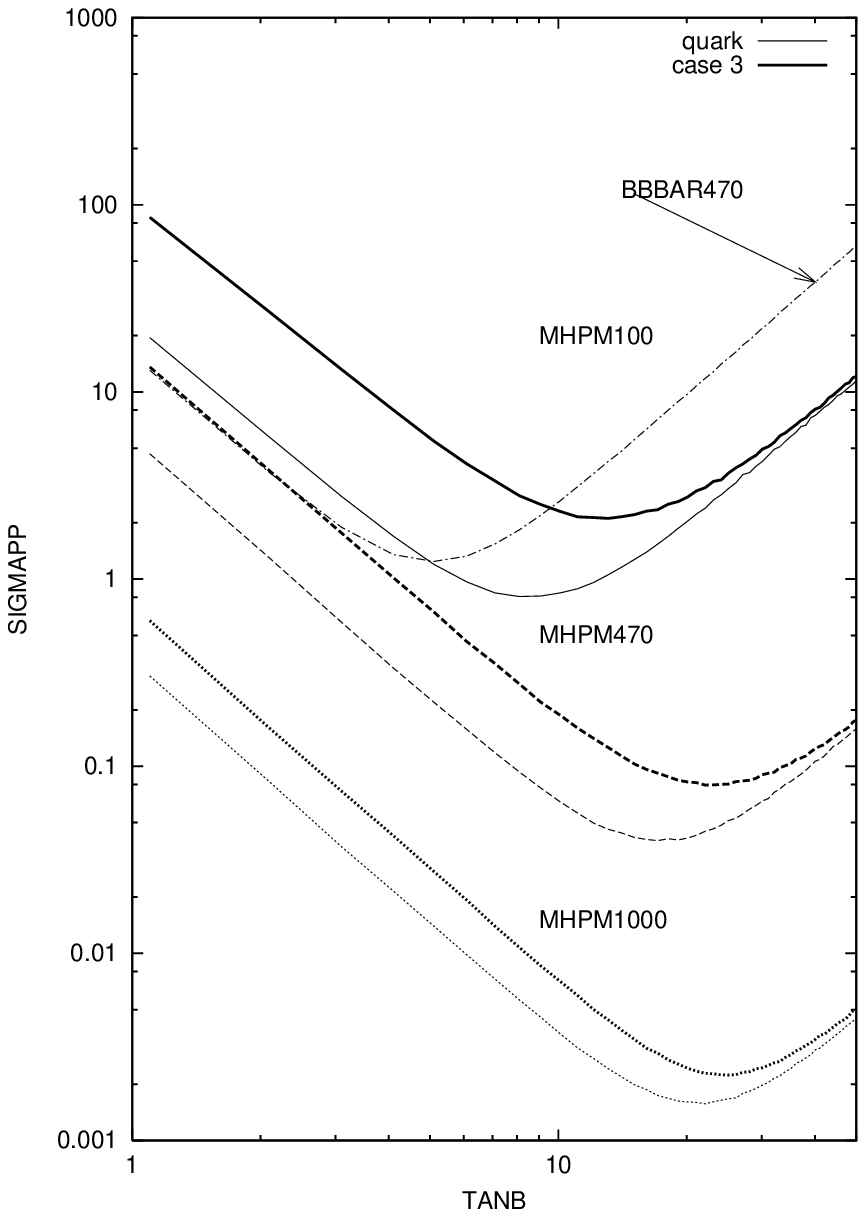}}
    \caption{\label{tb-case3} Hadronic cross section for $H^+ W^-$
        production via gluon fusion versus $\tan\beta$ for 
        three values of the charged-Higgs mass $\mhpm$ (100,470,1000 $\gev$). 
        The squark case 3 (thick lines) is compared to the case of 
        decoupling squarks (thin lines).
            }
  \end{center}
\end{figure}

The $\tan\beta$ dependence of the cross section in the large-mixing 
cases is shown in Figures~\ref{tb-case1} to \ref{tb-case3}.
The squark cases with mixing are all rather similar in their behavior.
Therefore we concentrate on  
Case~3 in Figure \ref{tb-case3}, as the most interesting example,
for a more explicit discussion. 
In Figure \ref{tb-case3},
the hadronic cross section is displayed for three values of 
$\mhpm$ (100,470, 1000 $\gev$) in Case 3 (thick lines) and 
it is compared to the case of decoupling squarks (thin lines).  
From the logarithmic plot one can read off the fact that the enhancement
due to squark effects in Case 3 with respect to the squark decoupling case 
keeps almost the same relative size for $\tan\beta$ values below
the position of the minimum of the hadronic cross section. 
With further increasing $\tan\beta$ the enhancement diminishes slowly. 
Taking the case of $\mhpm = 470\,\gev$ as an example, 
the magnitude of enhancement ranges from a factor of about 
3 in the range $1 \leq \tan\beta \leq 10$ to 1.1 at $\tan\beta = 50$.
In addition, there is a thin dot-dashed line in Figure \ref{tb-case3} 
showing the hadronic cross section which originates only from the 
$b\bar{b}$-annihilation subprocess for the most interesting 
$\mhpm$-value of 470 $\gev$, where the contribution by gluon fusion 
to the inclusive hadron process gets as important as the one 
by $b\bar{b}$-annihilation for $\tan\beta=1.5$
(see Figure \ref{mhpm-mixing}). 
It turns out that this special feature in the large mixing cases 
is valid for small values of $\tan\beta$.
The range of validity is limited from above roughly by the 
position of the minimum of the hadronic cross section originating
only from $b\bar{b}$-annihilation, which is at $\tan\beta \approx 5$ 
in the case considered here. This is because for $\tan\beta$ greater 
than 5, the $b\bar{b}$-annihilation cross section is again rising, while  
the gluon fusion cross section has not approached its minimum yet.

\bigskip
The figures discussed in this paper are based on calculations
assuming stable virtual particles, i.e. the finite widths of quarks
and squarks have been neglected. This treatment is 
a resonable approximation since the widths of the heavy  
quarks and squarks are small compared to their masses.
The widths of the squarks depend on the mass spectrum of the gauginos,
which determine the kinematically allowed squark-decay channels
but otherwise do not enter our calculation.
For example, the light sbottom, which plays the crucial role
in the threshold effects in the squark-box amplitudes,
has a width of the order of 
1 $\gev$, or much less if the decay channel $\Sbot_1
\to t \chi^-$ is kinematically closed (see e.g.~\cite{squark-widths}).

For a quantitative statement, 
we have estimated the finite-width effects for the squarks 
and the top quark in our calculation. 
Taking into account a finite squark width yields a
slight reduction of the threshold
peak in the hadronic cross section (Figure \ref{mhpm-mixing})
and no change off the peak. Specifically, we get a 
reduction of the peak value displayed in Figure \ref{mhpm-mixing} 
of 1 to 3~\% in the various scanarios
for squark widths of 0.1 $\gev$ and of about 10 \% for 1 $\gev$,
almost independent of $\tan\beta$.
The inclusion of the top quark width results in a reduction 
of 9~\% for $\tan\beta = 1.5$ and 18~\% for $\tan\beta = 6$
of the threshold peak located at $\mhpm = 210\;\gev$ in
Figure \ref{mhpm-mixing}.

\section{Conclusions}

We have discussed the charged-Higgs-boson production process 
associated with a $W$ boson at hadron colliders. 
The hadronic cross section from the subprocess $gg \to W^-H^+$ has 
been calculated in the MSSM and the squark-loop contributions 
examined in comparison with the quark-loop effects for various MSSM 
parameters, which are chosen respecting bounds from the 
electroweak precision measurement and the squark direct search.
We find that the squark (stop and sbottom) loop effects can be of 
about the same order as the $t$-$b$ loop effects.  
In addition, for the maximum mixing between $\tilde{t}_L$ and 
$\tilde{t}_R$ with the mixing angle $\theta_t \sim \pi/4$, 
the hadronic cross section from $gg\to W^\pm H^\mp$ is extensively 
enhanced by the threshold effects of $\tilde{t}_1$ and $\tilde{b}_{1,2}$, 
where $\tilde{t}_1$ is the lighter stop. 
Therefore, the hadronic cross section via gluon fusion 
can reach the size of the cross section via the tree 
$b\overline{b}$ annihilation subprocess for smaller $\tan\beta$.

\bigskip\bigskip
\noindent {\em Acknowledgement.} 

This work was supported in part by the Alexander von Humboldt 
Foundation and by the Deutsche Forschungsgemeinschaft.
Parts of the calculations have been performed on the QCM
cluster at the University of Karlsruhe, supported by the 
DFG-Forschergruppe ''Quantenfeldtheorie, Computeralgebra und 
Monte-Carlo-Simulation''.

\newpage

\appendix

\section*{Appendix}

\section{Squark masses and mixing}

Squarks are introduced as the super-partners of quarks, so that  
there are three generations of isospin-doublets and -singlets
corresponding to the quarks.
The scalar partners of the $L$- and $R$-chiral quarks, in general,
mix to form the mass eigenstates.
 
For the third generation of squarks, the mass-squared matrices 
in the $L$-$R$ basis have the form 
\begin{equation}\label{M-stop}
M_{\Stop}^2 =
\left(
\begin{array}[2]{cc}
M_{\tilde{Q}}^2+m_t^2+m_Z^2(\frac{1}{2}-e_t s_w^2)\cos2\beta &
m_t X_t \\
m_t X_t  &
M_{\tilde{U}}^2+m_t^2+m_Z^2 e_t s_w^2\cos2\beta\\
\end{array}
\right)         \ehsp,
\end{equation}
and
\begin{equation}\label{M-sbottom}
M_{\Sbot}^2 =
\left(
\begin{array}[2]{cc}
M_{\tilde{Q}}^2+m_b^2+m_Z^2(-\frac{1}{2}-e_b s_w^2)\cos2\beta &
m_b X_b \\
m_b X_b &
M_{\tilde{D}}^2+m_b^2+m_Z^2 e_b s_w^2\cos2\beta\\
\end{array}
\right)         \ehsp,
\end{equation}
where 
\begin{align}
  X_t &= A_t - \mu \cot\beta, \\
  X_b &= A_b - \mu \tan\beta. 
\end{align}
For squarks of the first two generation, the mass matrices are 
obtained analogously. 
In Eqs.~(\ref{M-stop}) and (\ref{M-sbottom}),  
the symbols $e_t$ and $e_b$ denote the electric charge of top- and
bottom-quarks; $\mu$ is the supersymmetric Higgs mass parameter,
$M_{\tilde{Q}}$ the soft-breaking mass parameter for the 
squark iso-doublet $(\Stop_L, \Sbot_L)$, and
$M_{\tilde{U}}$ and $M_{\tilde{D}}$ are the soft-breaking 
mass parameters for the iso-singlets $\Stop_R$ and $\Sbot_R$.
They can be different for each generation, but for simplicity
we will assume equal values for all generations in our numerical analysis.
$A_t$ and $A_b$ are the parameters of the soft-breaking scalar three-point 
interactions of top- and bottom-squarks with the Higgs fields.

We restrict our analysis to real parameters, so that the mass matrices 
for sfermions ($\tilde{f}_L$ and $\tilde{f}_R$) are real and can be 
diagonalized by introducing the mixing angles $\theta_{\tilde{f}}$. 
The mass eigenstates $\tilde{f}_1$ and $\tilde{f}_2$ are obtained 
as\begin{equation*}
\left(
\begin{array}[1]{c}
\tilde{f}_1\\
\tilde{f}_2
\end{array}
\right)
=
\left(
\begin{array}[2]{cc}
\cos\theta_{\tilde{f}} & -\sin\theta_{\tilde{f}}\\
\sin\theta_{\tilde{f}} &  \cos\theta_{\tilde{f}}
\end{array}
\right)
\left(
\begin{array}[1]{c}
\tilde{f}_L\\
\tilde{f}_R
\end{array}
\right) \, .
\end{equation*}
The off-diagonal 
parts in the mass-squared matrices~(\ref{M-stop}),(\ref{M-sbottom}) 
are proportional to the fermion masses. 
Therefore, the mixing effects are important mainly for third
generation squarks.

\section{Helicity amplitudes}\label{helamp}

In the analytic expressions for the helicity amplitudes scalar 3- 
and 4-point functions appear for which the following 
conventions and shorthand notations 
are introduced:
\begin{align*}
C_0^{lm,ABC}   & = \frac{1}{i \pi^2}
        \idvk \frac{1}{
        [k^2\!-\!m_A^2]
        [(k\!+\!p_l)^2\!-\!m_B^2]
        [(k\!+\!p_l\!+\!p_m)^2\!-\!m_C^2]} 
\; ,\\
D_0^{klm, ABCD} & = \frac{1}{i \pi^2}
        \idvk \frac{1}{
        [k^2\!-\!m_A^2]
        [(k\!+\!p_k)^2\!-\!m_B^2]
        [(k\!+\!p_k\!+\!p_l)^2\!-\!m_C^2]
        [(k\!+\!p_k\!+\!p_l\!+\!p_m)^2\!-\!m_D^2]} 
\; .
\end{align*}
Furthermore, in the box amplitudes
tensor coefficients appear, where the upper indices denote
momenta and masses in the same way as above. The lower index of the 
tensor coefficients gives the number in the tensor decomposition
and corresponds precisely to the naming in \cite{PV} except for
the fact that the Minkowski metric is used as implemented in
the Fortran package AAFF \cite{ff}, which has been used in the 
numerical evaluation. To match the definition of the loop
integrals the external momenta are all chosen as 
incoming. Thus we have the following translation to the definition 
of section \ref{partcs}:
\begin{align*}
p_1&=k & p_2 &=\bar{k} & p_3&=-p & p_4&=-\bar{p}
\; .
\end{align*}
In some $C$-functions a sum of two momenta $p_l+p_k$ appears as one of 
the momentum arguments, which is then denoted by $(lk)$.
The amplitude for $gg\to W^- H^+$ is divided into contributions from
quark and squark loops and further into triangle-type and box-type
diagrams:
\begin{align}
\MM_{\sigma\bar{\sigma}\lambda} & =
 \MM_{\sigma\bar{\sigma}\lambda}^{q,\triangle}
+ \MM_{\sigma\bar{\sigma}\lambda}^{q,\square}
+ \MM_{\sigma\bar{\sigma}\lambda}^{\Sq,\triangle}
+ \MM_{\sigma\bar{\sigma}\lambda}^{\Sq,\square}
\end{align}
Regarding the squark loops, we include the 2-point loops
in the triangle contribution and the triangles without Higgs 
exchange in the box contribution.

\subsection{Quark contributions}

\subsubsection*{Triangle contributions}

\begin{align}
 \MM_{\sigma\bar{\sigma}\lambda}^{q,\triangle} & =
( K_{1\, \sigma\bar{\sigma}\lambda} + K_{2\, \sigma\bar{\sigma}\lambda})
T_{1}^{q,\triangle}
+
( K_{10\, \sigma\bar{\sigma}\lambda} + K_{11\, \sigma\bar{\sigma}\lambda})
T_{2}^{q,\triangle}
\end{align}

\begin{multline*}
T^{q,\triangle}_1 = 
\frac{i\, g_S^{2}}{\pi^2}
\bigg[
g[H^0 H^\pm W^\mp] \times\\
\times
\bigg(
\frac{m_t\, g_s[H^0 t t]}{\hat{s}- {\it m_H}^{2}}
\Big( 
(
\frac {\hat{s}}{2} - 2\,{\it m_t}^{2}
) {\it C_{0}^{12,ttt}}
-1
\Big)
+ 
\frac{m_b\, g_s[H^0 b b]}{\hat{s} - {\it m_H}^{2}}
\Big(
(
\frac {\hat{s}}{2} - 2\,{\it m_b}^{2}
) {\it C_{0}^{12,bbb}}
-1
\Big)
\bigg)\\
+
g[h^0 H^\pm W^\mp]
\bigg(
\frac{m_t\,  g_s[h^0 t t]}{\hat{s}- {\it m_h}^{2}}
\Big( 
(
\frac {\hat{s}}{2} - 2\,{\it m_t}^{2}
) {\it C_{0}^{12,ttt}}
-1
\Big) 
+
\frac{m_b\, g_s[h^0 b b]}{\hat{s} - {\it m_h}^{2}}
\Big(
(
\frac {\hat{s}}{2} - 2\,{\it m_b}^{2}
) {\it C_{0}^{12,bbb}}
-1
\Big) 
\bigg)
\bigg]
\end{multline*}
\begin{multline*}
\lefteqn{
{\it T^{q,\triangle}_2} = 
\frac{g_S^{2}\, g[A^0 H^\pm W^\mp]}{\pi^2\,(\hat{s} - {\it m_A}^{2})}
\bigg( 
{\it m_t}\,{\it g_p[A^0 t t]}\, {\it C_{0}^{12,ttt}} 
+
{\it m_b}\, {\it g_p[A^0 b b]}\, {\it C_{0}^{12,bbb}}
\bigg)  
}
\end{multline*}
The universal helicity factors $K_{i\, \sigma\bar{\sigma}\lambda}$
are listed below in appendix \ref{k-factors}.

\subsubsection*{Box contributions}

\begin{align}
 \MM_{\sigma\bar{\sigma}\lambda}^{q,\square} & =
 \sum_{i=1}^{24} K_{i\, \sigma\bar{\sigma}\lambda}
        T_{i}^{q,\square}
\end{align}

\begin{align*}
T^{q,\square}_1 & = \frac{i\, g_S^{2}\, g_2}{4\sqrt{2} \pi^{2}}\,
\Big[
m_t\, g_{+} 
\Big(   
R_1
+ D_{12}^{132,bbtt}
+ D_{22}^{132,bbtt} 
 - D_{22}^{231,bbtt} 
- D_{13}^{124,tttb} 
- D_{13}^{214,tttb}
+ D_{23}^{123,bbbt} \\ 
& - D_{23}^{124,tttb}
+ D_{23}^{213,bbbt} 
- D_{23}^{214,tttb} 
  \Big)  
 + m_b\, g_{-} 
\Big(   
R_1
+ D_{0}^{132,bbtt} 
+ 4\,D_{12}^{132,bbtt} 
+ D_{12}^{231,bbtt}  \\
& + D_{22}^{231,bbtt}
+ 3\, D_{22}^{132,bbtt} 
+ D_{13}^{123,bbbt} 
+ D_{13}^{213,bbbt}
+ 3\, D_{23}^{123,bbbt} 
+ D_{23}^{124,tttb} 
+ 3\, D_{23}^{213,bbbt}
+ D_{23}^{214,tttb}
  \Big)  
\Big]
\\
T^{q,\square}_2 & = T^{q,\square}_1 (p_1 \leftrightarrow p_2)
\\
R_1 & = 2\,( D_{24}^{132,bbtt}
+ D_{26}^{231,bbtt} 
- D_{33}^{124,tttb} 
- D_{33}^{214,tttb}
+ D_{36}^{132,bbtt}
+ D_{37}^{123,bbbt}
+ D_{37}^{124,tttb}
+ D_{38}^{231,bbtt} \\
& + D_{39}^{213,bbbt} 
+ D_{39}^{214,tttb}
)
\\
T^{q,\square}_3 & = 
\frac{i\, g_S^{2}\, g_2 }{ 2\sqrt{2} \pi^{2} } \frac{1}{8}\,
\Big[
m_t\, g_{+}
\Big(   
( - \hat{u} + m_{W}^{2} )
D_{0}^{214,tttb} 
- D_{0}^{123,bbbt}\,\hat{s}
+ (\,\hat{t} - m_{H\pm}^{2}\,)\,
        ( D_{0}^{132,bbtt} - D_{0}^{124,tttb} )\\
& + 4( D_{27}^{132,bbtt} 
+  D_{27}^{213,bbbt}
-  D_{27}^{214,tttb}
-  D_{27}^{231,bbtt} 
+  D_{27}^{123,bbbt}
-  D_{27}^{124,tttb} ) \\
& + 2 \,( C_{0}^{34,tbt}
-   C_{0}^{34,btb} ) + R_2
  \Big) 
 + m_b\, g_{-} 
\Big(    
(m_{H\pm}^{2} - 2\,\hat{s} - \hat{t}\,)\,D_{0}^{123,bbbt} 
- D_{0}^{132,bbtt}\,\hat{s}\\
& - (\, - \hat{t} + \hat{s} + m_{H\pm}^{2}\,)\,D_{0}^{213,bbbt}
- D_{0}^{124,tttb}\,\hat{s}
- (\, - \hat{u} + m_{W}^{2}\,)\,D_{0}^{231,bbtt} \\
& +4 ( 3 D_{27}^{132,bbtt}
+ 3 D_{27}^{213,bbbt} 
+ D_{27}^{214,tttb}
+ D_{27}^{231,bbtt}
+ 3 D_{27}^{123,bbbt} 
+ D_{27}^{124,tttb}) \\
& - 2\,( C_{0}^{34,tbt} 
+ 3\,C_{0}^{34,btb}  )
+R_2
\Big)
\Big]
\\
T^{q,\square}_4 &= T^{q,\square}_3 (p_1 \leftrightarrow p_2)
\\
R_2 & =
\hat{s}( - D_{11}^{124,tttb} 
- D_{11}^{123,bbbt}
- D_{11}^{132,bbtt} 
- D_{12}^{213,bbbt} 
- D_{12}^{214,tttb}
+ D_{13}^{214,tttb}
+ D_{13}^{124,tttb} 
- D_{13}^{231,bbtt} ) \\
& + 8( D_{311}^{123,bbbt}
+ D_{311}^{132,bbtt} 
+ D_{311}^{124,tttb}
+ D_{312}^{214,tttb} 
+ D_{312}^{213,bbbt}
+ D_{313}^{231,bbtt}
- D_{313}^{214,tttb} 
- D_{313}^{124,tttb} )\\
& - 4( \,C_{11}^{(12)3,bbt}
+ \,C_{11}^{(12)4,ttb} 
- \,C_{12}^{(12)4,ttb} )
\\
T^{q,\square}_5 & =  
\frac{i\, g_S^{2}\, g_2}{2\sqrt{2} \pi^{2}} \frac{1}{8}
\Big[
m_t\, g_{+} 
\Big(    
\hat{s}\,(
- \, D_{0}^{214,tttb}
+  D_{0}^{124,tttb}
+  D_{12}^{132,bbtt}
+  D_{13}^{123,bbbt}
+  D_{13}^{213,bbbt}
)\\
& - (\,2\, m_{W}^{2} - \hat{s} - 2\,\hat{t}\,)
\,(  D_{13}^{124,tttb} 
+  D_{13}^{214,tttb} 
-  D_{12}^{231,bbtt} )
+R_3^+
  \Big) \\
& +  
m_b\,g_{-} 
\Big(   
\hat{s}\,(
   D_{0}^{123,bbbt}
-  D_{0}^{213,bbbt}
+  D_{0}^{231,bbtt}
-  D_{13}^{124,tttb}
-  D_{13}^{214,tttb}
-  D_{12}^{231,bbtt}
)\\
& + (\,2\, m_{W}^{2} - \hat{s} - 2\,\hat{t}\,) 
   \,(  D_{0}^{132,bbtt} 
   +  D_{12}^{132,bbtt} 
   +  D_{13}^{123,bbbt} 
   +  D_{13}^{213,bbbt}
)
+R_3^-
) 
  \Big) 
\Big]
\\
T^{q,\square}_6 & = T^{q,\square}_5 ( p_1 \leftrightarrow p_2 )
\end{align*}
\begin{align*}
R_3^\pm & =
8\,(
   D_{312}^{132,bbtt} 
+  D_{312}^{231,bbtt} 
+  D_{313}^{123,bbbt}
+  D_{313}^{213,bbbt}
-  D_{313}^{214,tttb} 
-  D_{313}^{124,tttb}
+  D_{27}^{132,bbtt}
)\\
& -2\,(
  2\, C_{12}^{2(13),bbt}
-  C_{12}^{24,ttb}
+  C_{0}^{24,ttb} 
+  C_{11}^{(13)2,btt}
) 
\mp 2\,(        
         C_{0}^{23,ttb} 
        +  C_{12}^{2(14),ttb} 
        +  C_{11}^{32,btt}
)
\\
T^{q,\square}_7 & = 
\frac{ g_S^{2}\, g_2}{2\sqrt{2} \pi^{2}} \frac{1}{4}
\Big[
m_t\,g_{+} 
\Big( 
R_{4}   
+  D_{13}^{132,bbtt}
  \Big) \\
& +
m_b\,g_{-} 
\Big(
R_{5}    
-  D_{0}^{123,bbbt} 
+  D_{0}^{213,bbbt}
-  D_{11}^{123,bbbt} 
+  D_{11}^{132,bbtt}
+  D_{11}^{213,bbbt}
   \Big) 
\Big]
\\
T^{q,\square}_8 & = 
\frac{ g_S^{2}\, g_2}{2\sqrt{2} \pi^{2}} \frac{1}{4}
\Big[
m_t\,g_{+} 
\Big(   
-R_{4} (p_1 \leftrightarrow p_2 )
-  D_{13}^{132,bbtt}
  \Big)
 +
m_b\,g_{-} 
\Big(
-R_{5} (p_1 \leftrightarrow p_2 )    
-  D_{11}^{231,bbtt}
  \Big)  
\Big]
\\
R_{4} & = 
-  D_{0}^{214,tttb} 
-  D_{11}^{214,tttb}
-  D_{12}^{124,tttb}
-  D_{12}^{231,bbtt} 
-  D_{13}^{123,bbbt}
+  D_{13}^{124,tttb} \\
R_{5} & = 
    D_{0}^{132,bbtt}
 +  D_{11}^{231,bbtt} 
 -  D_{12}^{231,bbtt}
 +  D_{13}^{124,tttb}
\\
T^{q,\square}_9 & = 
\frac{ g_S^{2}\, g_2}{2\sqrt{2} \pi^{2}} \frac{1}{4}
\Big[
m_t\,g_{+} 
\Big(    
- \,\, D_{11}^{124,tttb} 
+ \, D_{11}^{214,tttb}
+ \, D_{12}^{124,tttb}
- \, D_{12}^{214,tttb} 
  \Big) \\
& + m_b\,g_{-}
\Big(   
\, D_{11}^{123,bbbt} 
- \, D_{11}^{213,bbbt}
- \, D_{13}^{123,bbbt}
+ \, D_{13}^{213,bbbt} 
  \Big)
\Big]
\\
T^{q,\square}_{10} & = 
\frac{ g_S^{2}\, g_2}{2\sqrt{2} \pi^{2}} \frac{1}{4}
\Big[
m_t\,g_{+} 
\Big(
R_{6}    
+  D_{12}^{124,tttb}
-  D_{12}^{214,tttb}
 +  D_{13}^{214,tttb}
+  D_{23}^{124,tttb}
+  D_{24}^{124,tttb} 
-  D_{25}^{124,tttb} \\
& 
 -  D_{26}^{124,tttb} 
  \Big)
+
m_b\,g_{-} 
\Big(
  R_{7}
+ R_{8}   
  \Big)  
\Big]
\\
T^{q,\square}_{11} & = 
\frac{ g_S^{2}\, g_2}{2\sqrt{2} \pi^{2}} \frac{1}{4}
\Big[
m_t\,g_{+}
\Big(
R_{6} (p_1 \leftrightarrow p_2 )         
+  D_{13}^{214,tttb}
+  D_{22}^{124,tttb}
+  D_{23}^{124,tttb} 
- 2\, D_{26}^{124,tttb} 
  \Big) \\
& +
m_b\,g_{-} 
\Big(   
R_{7} (p_1 \leftrightarrow p_2 )    
+R_{8} 
  \Big)  
\Big] \\
R_{6} & =
-  D_{11}^{124,tttb}
+  D_{11}^{132,bbtt}
+  D_{12}^{123,bbbt} 
+  D_{12}^{213,bbbt}
+  D_{21}^{132,bbtt}
+  D_{22}^{213,bbbt} 
+  D_{24}^{123,bbbt}
+  D_{25}^{231,bbtt}
\\
R_{7} & =
   D_{0}^{123,bbbt} 
+  D_{0}^{132,bbtt}
+  D_{0}^{213,bbbt}
+ 2\, D_{11}^{123,bbbt} 
+  D_{11}^{132,bbtt}
+  D_{12}^{124,tttb}
+ 2\, D_{12}^{213,bbbt} 
+  D_{12}^{214,tttb}  \\
& +  D_{21}^{123,bbbt} 
+  D_{22}^{214,tttb}
+  D_{23}^{124,tttb}
+  D_{23}^{214,tttb} 
+  D_{23}^{231,bbtt}
+  D_{24}^{124,tttb}
+  D_{24}^{213,bbbt} 
-  D_{25}^{124,tttb} \\
& +  D_{25}^{132,bbtt}
-  D_{26}^{124,tttb} 
- 2\, D_{26}^{214,tttb} 
\\
R_{8} & =
-  D_{13}^{124,tttb}
+  D_{13}^{132,bbtt} 
-  D_{13}^{214,tttb}
+  D_{13}^{231,bbtt}
\\
T^{q,\square}_{12} & = 
\frac{ g_S^{2}\, g_2}{2\sqrt{2} \pi^{2}} \frac{1}{4}
\Big[
m_t\,g_{+}
 \Big(   
  R_{9}
+ R_{10}
+   D_{0}^{124,tttb} 
-  D_{0}^{214,tttb} 
+  D_{11}^{124,tttb}
-  D_{11}^{214,tttb} 
  \Big) \\
& +
m_b\,g_{-}
\Big(   
R_{11}
  \Big) 
\Big]
\\
T^{q,\square}_{13} & = 
\frac{ g_S^{2}\, g_2}{2\sqrt{2} \pi^{2}} \frac{1}{4}
\Big[
m_t\,g_{+} 
\Big(   
- R_{9} (p_1 \leftrightarrow p_2 )
+ R_{10}     
-  D_{23}^{124,tttb} 
- 2\, D_{23}^{214,tttb}
+ 2\, D_{26}^{214,tttb} 
  \Big)  \\
& +
m_b\,g_{-} 
\Big( 
- R_{11} (p_1 \leftrightarrow p_2 )   
  \Big) 
\Big]
\\
R_{9} & = 
-  D_{12}^{124,tttb}
-  D_{12}^{132,bbtt}
+  D_{13}^{124,tttb}
+ 2 D_{13}^{132,bbtt} 
+  D_{13}^{213,bbbt}
+ 2 D_{25}^{213,bbbt} 
- 2 D_{25}^{214,tttb}
+ 2 D_{26}^{132,bbtt}
\\
R_{10} & = 
   D_{23}^{124,tttb} 
+ 2 D_{23}^{214,tttb} 
-  D_{24}^{132,bbtt}
+  D_{24}^{231,bbtt} 
+  D_{26}^{123,bbbt} 
-  D_{26}^{124,tttb}
-  D_{26}^{213,bbbt} 
\\
R_{11} & = 
   D_{0}^{132,bbtt} 
+  D_{0}^{213,bbbt} 
+  D_{11}^{132,bbtt}
+  D_{11}^{213,bbbt} 
+  D_{11}^{231,bbtt}
+  D_{12}^{231,bbtt}
+  D_{13}^{123,bbbt} \\
&
+  D_{13}^{132,bbtt} 
+  D_{13}^{213,bbbt}
-  D_{13}^{214,tttb} 
-  D_{13}^{231,bbtt}
+  D_{23}^{124,tttb}
+  D_{23}^{214,tttb} 
+ 2 D_{24}^{231,bbtt}
-  D_{25}^{123,bbbt} \\
&
+  D_{25}^{213,bbbt} 
- 2 D_{25}^{214,tttb} 
+ 2 D_{26}^{123,bbbt}
-  D_{26}^{124,tttb} 
+  D_{26}^{132,bbtt} 
+  D_{26}^{214,tttb}
-  D_{26}^{231,bbtt} 
\end{align*}
\begin{align*}
T^{q,\square}_{14} &= 
\frac{ g_S^{2}\, g_2}{2\sqrt{2} \pi^{2}} \frac{1}{4}
\Big[
m_t\,g_{+}
\Big(    
-  D_{0}^{124,tttb} 
-  D_{11}^{124,tttb} 
+  D_{11}^{132,bbtt}
-  D_{12}^{214,tttb} 
+  D_{13}^{123,bbbt} 
+  D_{13}^{214,tttb}\\
& +  D_{21}^{132,bbtt}
-  D_{22}^{214,tttb}
+  D_{23}^{124,tttb}
-  D_{23}^{214,tttb} 
+  D_{25}^{123,bbbt} 
-  D_{25}^{124,tttb}
+ 2\, D_{26}^{214,tttb} 
+  D_{26}^{231,bbtt} 
  \Big)  \\
&
+
m_b\,g_{-} 
\Big(
   D_{0}^{123,bbbt}
+  D_{0}^{132,bbtt} 
+ 2\, D_{11}^{123,bbbt}
+  D_{11}^{132,bbtt} 
+  D_{12}^{132,bbtt} 
+  D_{12}^{213,bbbt}
-  D_{13}^{124,tttb} \\
&
+  D_{21}^{123,bbbt} 
+  D_{23}^{124,tttb}
+  D_{24}^{132,bbtt} 
-  D_{25}^{124,tttb} 
+  D_{26}^{213,bbbt}
  \Big) 
\Big]
\\
T^{q,\square}_{15} &= 
\frac{ g_S^{2}\, g_2}{2\sqrt{2} \pi^{2}} \frac{1}{4}
\Big[
m_t\,g_{+}
\Big(   
   D_{11}^{124,tttb} 
-  D_{12}^{124,tttb} 
-  D_{12}^{214,tttb}
+  D_{12}^{231,bbtt} 
+  D_{13}^{123,bbbt} 
+  D_{13}^{214,tttb}
+  D_{23}^{124,tttb} \\
&
-  D_{23}^{214,tttb} 
-  D_{24}^{214,tttb}
+  D_{24}^{231,bbtt} 
+  D_{25}^{132,bbtt} 
+  D_{25}^{214,tttb}
+  D_{26}^{123,bbbt} 
-  D_{26}^{124,tttb} 
+  D_{26}^{214,tttb}
  \Big) \\
&
+
m_b\,g_{-} 
\Big(  
   D_{0}^{123,bbbt} 
+  D_{11}^{123,bbbt} 
+  D_{11}^{213,bbbt} 
+  D_{12}^{123,bbbt}
-  D_{13}^{124,tttb} 
+  D_{13}^{132,bbtt} \\
&
+  D_{23}^{124,tttb}
+  D_{24}^{123,bbbt} 
+  D_{25}^{213,bbbt} 
-  D_{26}^{124,tttb}
+  D_{26}^{132,bbtt} 
  \Big) 
\Big]
\\
T^{q,\square}_{16} &= 
\frac{ g_S^{2}\, g_2}{2\sqrt{2} \pi^{2}} \frac{1}{4}
\Big[
m_t\,g_{+}
\Big(   
   D_{12}^{132,bbtt}
+  D_{13}^{124,tttb}
+ 2\, D_{13}^{214,tttb} 
+ 2\, D_{22}^{132,bbtt} 
+  D_{22}^{231,bbtt}
+  D_{23}^{123,bbbt} \\
&
+  D_{23}^{124,tttb} 
+ 2\, D_{23}^{213,bbbt}
+ 3\, D_{23}^{214,tttb} 
-  D_{24}^{132,bbtt} 
-  D_{26}^{214,tttb}
   \Big) \\
&
+
m_b\,g_{-}
\Big(    
   D_{0}^{132,bbtt} 
+ 2\, D_{12}^{132,bbtt} 
+ 2\, D_{12}^{231,bbtt}
+  D_{13}^{123,bbbt}
+  D_{13}^{213,bbbt} 
+  D_{22}^{132,bbtt} 
+ 2\, D_{22}^{231,bbtt} \\
&
+ 2\, D_{23}^{123,bbbt} 
+  D_{23}^{124,tttb} 
+  D_{23}^{213,bbbt}
+ 2\, D_{23}^{214,tttb} 
-  D_{25}^{123,bbbt} 
  \Big)  
\Big]
\\
T^{q,\square}_{17} &= 
\frac{ g_S^{2}\, g_2}{2\sqrt{2} \pi^{2}} \frac{1}{4}
\Big[
m_t\,g_{+}
\Big(   
   D_{12}^{132,bbtt}
+  D_{13}^{124,tttb}
+  D_{22}^{231,bbtt} 
+  D_{23}^{123,bbbt}
+  D_{23}^{124,tttb}
-  D_{23}^{214,tttb}\\
& 
+  D_{24}^{132,bbtt} 
+  D_{26}^{214,tttb}
   \Big)
+
m_b\,g_{-}
\Big(  
   D_{0}^{132,bbtt} 
+ 2\, D_{12}^{132,bbtt} 
+  D_{13}^{123,bbbt} 
+  D_{13}^{213,bbbt}
+  D_{22}^{132,bbtt}\\
& 
+  D_{23}^{124,tttb} 
+  D_{23}^{213,bbbt}
+  D_{25}^{123,bbbt} 
  \Big) 
\Big]
\\
T^{q,\square}_{18} &=
\frac{ g_S^{2}\, g_2}{2\sqrt{2} \pi^{2}} \frac{1}{8}
\Big[
m_t\,g_{+}
\Big(    
- \,{\it m_b}\,{\it m_t}\,
(
   D_{0}^{123,bbbt}
+  D_{0}^{213,bbbt} 
+2\, D_{0}^{132,bbtt}
+2\, D_{0}^{124,tttb}
+  D_{11}^{123,bbbt}
)\\
&
+ (\,{\it m_b}^{2} - {\it m_t}^{2} +  m_{H\pm}^{2}\,)\,
(
   D_{0}^{132,bbtt} 
 + D_{11}^{132,bbtt}
 - D_{12}^{214,tttb}
 + D_{13}^{214,tttb}
)
+ (\,\hat{u} -  m_{H\pm}^{2}\,)\, D_{13}^{231,bbtt} \\
&
+{\it m_b}^{2}\,
(
2\, D_{0}^{123,bbbt}
- 2\, D_{0}^{231,bbtt}
+ D_{0}^{124,tttb}
)
- ({\it m_t}^2 -  m_{H\pm}^2)
 D_{0}^{124,tttb} \\
&
+
(
 C_{11}^{(12)4,ttb} 
-  C_{12}^{(12)4,ttb} 
-  C_{0}^{13,ttb} 
+ 2\, C_{0}^{24,ttb} 
-  C_{0}^{12,ttt} 
-  C_{12}^{21,ttt}\\
& 
- 4\, D_{27}^{132,bbtt} 
- 4\, D_{27}^{213,bbbt} 
+ 2\, D_{27}^{123,bbbt}
+ 2\, D_{27}^{231,bbtt}
+ R_9 )
  \Big) \\
&
+ 
m_b\,g_{-}
\Big(   
(\,\hat{u} + {\it m_b}^{2}\,)
\,(
   D_{0}^{123,bbbt} 
+  D_{11}^{123,bbbt}
)
- (\,\hat{u} -  m_{H\pm}^{2}\,)
\,(
  D_{0}^{132,bbtt} 
+ D_{11}^{132,bbtt}
)\\
&
- (\,{\it m_b}^2 -  m_{H\pm}^2\,)\, D_{0}^{213,bbbt}
+ (\,\hat{t} -  m_{H\pm}^{2}\,)\, D_{13}^{231,bbtt}\\
&
+
(
 C_{0}^{14,ttb}
+  C_{0}^{24,ttb} 
-  C_{11}^{12,bbb} 
-  C_{12}^{(23)1,btt}
- 2\, C_{0}^{12,bbb} 
+  C_{12}^{31,btt} \\
&
+  C_{0}^{24,bbt}
-  C_{0}^{13,bbt}
+ 2\, D_{27}^{132,bbtt} 
+ 2\, D_{27}^{213,bbbt}
- 4\, D_{27}^{231,bbtt} 
- 4\, D_{27}^{123,bbbt}
+ R_9 )
  \Big) \\
&
 +
2\,{\it m_t}^{2}\,{\it m_b}\,
g_s[H^+\:\text{out}, t\:\text{in}, b\:\text{out}] \,
\Big(   
  D_{0}^{123,bbbt} 
+ 2\, D_{0}^{124,tttb}
+ 2\, D_{0}^{132,bbtt} 
+  D_{0}^{213,bbbt}
+  D_{11}^{123,bbbt} 
  \Big) 
\Big]
\\
R_9 &=
-  C_{12}^{14,ttb}
+  C_{12}^{24,ttb}
+  C_{11}^{(13)2,btt}
-  C_{12}^{2(13),bbt}
+  C_{11}^{14,ttb}
- 4\, D_{27}^{214,tttb}
+ 2\, D_{27}^{124,tttb}
\end{align*}
\begin{align*}
T^{q,\square}_{19} &= 
\frac{ g_S^{2}\, g_2}{2\sqrt{2} \pi^{2}} \frac{1}{4}
\Big[
m_t\,g_{+}
\Big(  
- \, D_{0}^{124,tttb} 
-  D_{11}^{124,tttb}
-  D_{11}^{214,tttb}
-  D_{12}^{132,bbtt} 
+  D_{13}^{132,bbtt} 
-  D_{13}^{213,bbbt}\\
&
+  D_{13}^{214,tttb} 
-  D_{23}^{214,tttb} 
-  D_{24}^{132,bbtt}
-  D_{24}^{214,tttb} 
+  D_{25}^{132,bbtt} 
+  D_{25}^{214,tttb}
-  D_{26}^{213,bbbt} 
+  D_{26}^{214,tttb}   
\Big)\\
&
+
m_b\,g_{-} 
\Big(  
- \, D_{11}^{123,bbbt} 
+  D_{12}^{123,bbbt}
+  D_{13}^{214,tttb} 
-  D_{13}^{231,bbtt}\\
&
-  D_{23}^{214,tttb}
+  D_{24}^{123,bbbt} 
-  D_{25}^{123,bbbt}
+  D_{26}^{214,tttb}
-  D_{26}^{231,bbtt} 
  \Big)
\Big]
\\
T^{q,\square}_{20} &= 
\frac{ g_S^{2}\, g_2}{2\sqrt{2} \pi^{2}} \frac{1}{4}
\Big[
m_t\,g_{+}
\Big(   
- \, D_{11}^{214,tttb}
-  D_{13}^{213,bbbt}
+  D_{13}^{214,tttb} 
-  D_{21}^{214,tttb}
+  D_{23}^{132,bbtt}
-  D_{23}^{214,tttb} \\
&
-  D_{25}^{213,bbbt} 
+ 2\, D_{25}^{214,tttb}
-  D_{26}^{132,bbtt} 
\Big)
+ 
m_b\,g_{-} 
\Big(  
- \, D_{0}^{231,bbtt} 
-  D_{11}^{231,bbtt} 
-  D_{12}^{231,bbtt} \\
&
 +  D_{13}^{214,tttb} 
+  D_{22}^{123,bbbt}
-  D_{23}^{214,tttb}
-  D_{24}^{231,bbtt} 
+  D_{25}^{214,tttb}
-  D_{26}^{123,bbbt}
\Big)
\Big]
\\
T^{q,\square}_{21} &=
\frac{ g_S^{2}\, g_2}{2\sqrt{2} \pi^{2}} \frac{1}{4}
\Big[
m_t\,g_{+}
\Big(
 D_{22}^{132,bbtt}
+  D_{23}^{213,bbbt}
+  D_{23}^{214,tttb} 
-  D_{25}^{214,tttb}
-  D_{26}^{132,bbtt}
\Big)\\
&
+ 
m_b\,g_{-} 
\Big(  
 D_{0}^{231,bbtt} 
+ 2\, D_{12}^{231,bbtt} 
+  D_{22}^{231,bbtt}
+  D_{23}^{123,bbbt}
+  D_{23}^{214,tttb} 
-  D_{26}^{123,bbbt} 
\Big)
\Big]
\\
T^{q,\square}_{22} &= 
\frac{ g_S^{2}\, g_2}{2\sqrt{2} \pi^{2}} \frac{1}{4}
\Big[
m_t\,g_{+} 
\Big(  
2\, D_{12}^{231,bbtt}
+ 2\, D_{13}^{124,tttb}
+ 2\, D_{13}^{214,tttb} 
+  D_{22}^{132,bbtt}
+ 2\, D_{22}^{231,bbtt}
+ 2\, D_{23}^{123,bbbt} \\
&
+ 2\, D_{23}^{124,tttb}
+  D_{23}^{213,bbbt}
+  D_{23}^{214,tttb} 
+  D_{25}^{214,tttb}
+  D_{26}^{132,bbtt}
  \Big)\\
&
+ 
m_b\,g_{-} 
\Big(   
 D_{0}^{231,bbtt} 
+ 2\, D_{12}^{132,bbtt} 
+ 2\, D_{12}^{231,bbtt} 
+ 2\, D_{13}^{123,bbbt}
+ 2\, D_{13}^{213,bbbt} 
+ 2\, D_{22}^{132,bbtt} \\
&
+  D_{22}^{231,bbtt}
+  D_{23}^{123,bbbt} 
+ 2\, D_{23}^{124,tttb}
+ 2\, D_{23}^{213,bbbt}
+  D_{23}^{214,tttb} 
+  D_{26}^{123,bbbt} 
\Big)
\Big]
\\
T^{q,\square}_{23} &=
\frac{ g_S^{2}\, g_2}{2\sqrt{2} \pi^{2}} \frac{1}{8}
\Big[  
m_t\,g_{+} 
\Big(
  2\,{\it m_b}^{2}\,(
   D_{0}^{132,bbtt} 
+  D_{0}^{213,bbbt}
)
+ {\it m_b}\,{\it m_t}\,(
  2\, D_{0}^{231,bbtt}
+ 2\, D_{0}^{214,tttb}
-  D_{12}^{123,bbbt}
)\\
&
+ ( \hat{u} -  m_{H\pm}^{2}\,)
\,(
   D_{0}^{231,bbtt}
+  D_{11}^{231,bbtt}
)
+ (\,{\it m_b}^{2} - {\it m_t}^{2} +  m_{H\pm}^{2}\,)
\,(
   D_{13}^{132,bbtt}
+  D_{13}^{214,tttb}
-  D_{11}^{214,tttb} 
)\\
&
+ (
  4\, D_{27}^{123,bbbt}
- 2\, D_{27}^{213,bbbt}
+ 4\, D_{27}^{214,tttb}
+ 4\, D_{27}^{231,bbtt}
- 2\, D_{27}^{132,bbtt}\\
&
+ C_{11}^{(12)4,ttb} 
-  C_{12}^{(12)4,ttb} 
-  C_{11}^{21,ttt}
-  C_{0}^{24,bbt} 
-  C_{11}^{2(13),bbt}
+ R_{10} )
\Big)\\
&
+m_b\,g_{-}
\Big( 
2\,{\it m_b}^{2}\, D_{0}^{123,bbbt}
+ ( \hat{u} -  m_{H\pm}^{2}\,)
\,(
   D_{0}^{123,bbbt}
-  D_{13}^{132,bbtt}
) \\
& + 
(\,\hat{t} -  m_{H\pm}^{2}\,)
\,(
   D_{0}^{231,bbtt}
+  D_{11}^{231,bbtt}
)
+ (\,\hat{u} + {\it m_b}^{2}\,)\, D_{12}^{123,bbbt}\\
& + (
  4\, D_{27}^{132,bbtt}
- 2\, D_{27}^{214,tttb}
- 2\, D_{27}^{231,bbtt}
- 2\, D_{27}^{123,bbbt}
+ 4\, D_{27}^{213,bbbt}\\
&
- C_{11}^{(23)1,btt}
-  C_{0}^{24,ttb} 
+  C_{0}^{24,bbt}
-  C_{12}^{32,btt} 
+  C_{0}^{23,bbt} 
-  C_{12}^{12,bbb}
+  C_{11}^{2(13),bbt}
+R_{10})
\Big)\\
&
+
2{\it m_t}^{2}\,{\it m_b}\,
g_s[H^+\:\text{out}, t\:\text{in}, b\:\text{out}]\,
\Big(  
   D_{12}^{123,bbbt} 
- 2\, D_{0}^{231,bbtt} 
- 2\, D_{0}^{214,tttb}
\Big)
\Big]
\\
R_{10} &=
4\, D_{27}^{124,tttb}
-  C_{0}^{14,ttb}
-  C_{12}^{14,ttb}
+  C_{12}^{24,ttb}
-  C_{11}^{24,ttb}
+  C_{12}^{(13)2,btt}
\end{align*}
\begin{align*}
T^{q,\square}_{24} &=
\frac{ g_S^{2}\, g_2}{2\sqrt{2} \pi^{2}} \frac{1}{8}
\Big[
m_t\,g_{+} 
\Big(  
2\,{\it m_b}^{2}\, D_{0}^{132,bbtt} 
- 2\,{\it m_t}^{2}\, D_{0}^{214,tttb}
+ (\,{\it m_b}^{2} - {\it m_t}^{2} +  m_{H\pm}^{2}\,)\, D_{12}^{132,bbtt} \\
&
+ (\,\hat{u} -  m_{H\pm}^{2}\,)\, D_{12}^{231,bbtt} 
+ (\,{\it m_b}^{2} - {\it m_t}^{2} -  m_{H\pm}^{2}\,)\, D_{13}^{214,tttb}
- \,{\it m_t}\,{\it m_b}\, D_{13}^{123,bbbt} \\
&
- (
   C_{0}^{34,tbt}
- 2\, C_{0}^{24,ttb} 
+  C_{12}^{(12)4,ttb}
+  C_{12}^{2(13),bbt} 
- 6\, D_{27}^{214,tttb} 
+ 6\, D_{27}^{132,bbtt}
+ R_{11} ) 
\Big)\\
&
+ 
m_b\,g_{-} 
\Big(
2\,{\it m_b}^{2}\, D_{0}^{123,bbbt}
- (\,\hat{u} -  m_{H\pm}^{2}\,)
\,(
   D_{0}^{132,bbtt}
+  D_{12}^{132,bbtt} 
)\\
&
+ (\,\hat{t} -  m_{H\pm}^{2}\,)
\,(
   D_{0}^{231,bbtt}
+  D_{12}^{231,bbtt}
)
+ (\,\hat{u} + {\it m_b}^{2}\,)\, D_{13}^{123,bbbt}
- (
 C_{11}^{(23)1,btt}
-  C_{0}^{34,btb} 
-  C_{0}^{13,ttb} \\
&
-  C_{11}^{31,btt}
+  C_{0}^{23,ttb} 
-  C_{0}^{24,ttb} 
+  C_{11}^{32,btt}
-  C_{12}^{2(13),bbt} 
-  C_{0}^{24,bbt}
+ 6\, D_{27}^{123,bbbt}
+ R_{11} ) 
\Big)\\
&
+ 2\,{\it m_t}^{2}\,{\it m_b}\,
g_s[H^+\:\text{out}, t\:\text{in}, b\:\text{out}]
\, D_{13}^{123,bbbt} 
\Big]
\\
R_{11} &=
  C_{0}^{14,ttb}
-  C_{11}^{(13)2,btt}
+  C_{12}^{14,ttb}
-  C_{12}^{24,ttb}
\end{align*}
\noindent Here the shorthand
\begin{align*}
g_{\pm} & = g_s[H^+\:\text{out}, t\:\text{in}, b\:\text{out}]
        \pm g_p[H^+\:\text{out}, t\:\text{in}, b\:\text{out}]
\end{align*}
has been used.

\subsection{Squark contributions}

\subsubsection*{Triangle contributions}

\begin{align}
 \MM_{\sigma\bar{\sigma}\lambda}^{\Sq,\triangle} & =
( K_{1\, \sigma\bar{\sigma}\lambda} + K_{2\, \sigma\bar{\sigma}\lambda})
        T_{1}^{\Sq,\triangle}
\end{align}
\begin{multline*}
T^{\Sq,\triangle}_1 = 
\sum_{i=1}^{2}
\frac{i\, g_S^{2} }{4\, \pi^2}
\bigg[
\frac{ g[H^0 H^\pm W^\mp] }{ \hat{s} - {\it m_H}^{2} } \times\\
\times
\bigg(
g[ H^0 \Stop_i \Stop_i ] 
\Big(
1 + 2\,m_{\Stop_{i}}^{2}\,C_{0}^{12,\Stop_{i}\Stop_{i}\Stop_{i}} 
\Big)
+ {\it g[ H^0 \Sbot_i \Sbot_i ]}
\Big(
1 + 2\,m_{\Sbot_{i}}^{2}\,C_{0}^{12,\Sbot_{i}\Sbot_{i}\Sbot_{i}}
\Big)\\
+ \frac{ g[h^0 H^\pm W^\mp] }{ \hat{s} - {\it m_h}^{2} }
\bigg(
+ g[ h^0 \Stop_i \Stop_i ]
\Big(
1 + 2\,m_{\Stop_{i}}^{2}\,C_{0}^{12,\Stop_{i}\Stop_{i}\Stop_{i}}
\Big)
+ g[ h^0 \Sbot_i \Sbot_i ]
\Big(
1 + 2\,m_{\Sbot_{i}}^{2}\,C_{0}^{12,\Sbot_{i}\Sbot_{i}\Sbot_{i}}
\Big)
\bigg)
\bigg]
\end{multline*}

\subsubsection*{Box contributions}

\begin{align}
 \MM_{\sigma\bar{\sigma}\lambda}^{\Sq,\square} & =
 \sum_{i=1}^{6} K_{i\, \sigma\bar{\sigma}\lambda}
        T_{i}^{\Sq,\square}
\end{align}

\begin{align*}
T^{\Sq,\square}_1 &= 
\sum_{i,j=1}^{2}
\frac {i\, g_S^{2} }{2\, \pi^2} (
     D_{12}^{132,\Sbot_{i} \Sbot_{i} \Stop_{j} \Stop_{j}}
   - D_{12}^{132,\Stop_{j} \Stop_{j} \Sbot_{i} \Sbot_{i}} 
   + D_{22}^{132,\Sbot_{i} \Sbot_{i} \Stop_{j} \Stop_{j}} 
   - D_{22}^{132,\Stop_{j} \Stop_{j} \Sbot_{i} \Sbot_{i}} 
   + D_{24}^{132,\Sbot_{i} \Sbot_{i} \Stop_{j} \Stop_{j}} \\
&
   - D_{24}^{132,\Stop_{j} \Stop_{j} \Sbot_{i} \Sbot_{i}} 
   + D_{36}^{132,\Sbot_{i} \Sbot_{i} \Stop_{j} \Stop_{j}} 
   - D_{36}^{132,\Stop_{j} \Stop_{j} \Sbot_{i} \Sbot_{i}} 
   + D_{23}^{123,\Sbot_{i} \Sbot_{i} \Sbot_{i} \Stop_{j}} 
   - D_{23}^{123,\Stop_{j} \Stop_{j} \Stop_{j} \Sbot_{i}} 
   + D_{33}^{124,\Sbot_{i} \Sbot_{i} \Sbot_{i} \Stop_{j}} \\
&
   - D_{33}^{124,\Stop_{j} \Stop_{j} \Stop_{j} \Sbot_{i}}  
   + D_{37}^{123,\Sbot_{i} \Sbot_{i} \Sbot_{i} \Stop_{j}}
   - D_{37}^{123,\Stop_{j} \Stop_{j} \Stop_{j} \Sbot_{i}} 
   - D_{37}^{124,\Sbot_{i} \Sbot_{i} \Sbot_{i} \Stop_{j}}
   + D_{37}^{124,\Stop_{j} \Stop_{j} \Stop_{j} \Sbot_{i}})
\, g[\Stop_j \Sbot_i W^\pm]\, g[ H^\pm \Stop_j \Sbot_i ]
\\
T^{\Sq,\square}_2 &= 
\sum_{i,j=1}^{2}
\frac {i\, g_S^{2} }{2\, \pi^2}
(
     D_{26}^{132,\Sbot_{i} \Sbot_{i} \Stop_{j} \Stop_{j}}
   - D_{26}^{132,\Stop_{j} \Stop_{j} \Sbot_{i} \Sbot_{i}} 
   + D_{38}^{132,\Sbot_{i} \Sbot_{i} \Stop_{j} \Stop_{j}} 
   - D_{38}^{132,\Stop_{j} \Stop_{j} \Sbot_{i} \Sbot_{i}} 
   + D_{23}^{123,\Sbot_{i} \Sbot_{i} \Sbot_{i} \Stop_{j}} \\
&
   - D_{23}^{123,\Stop_{j} \Stop_{j} \Stop_{j} \Sbot_{i}}
   + D_{33}^{124,\Sbot_{i} \Sbot_{i} \Sbot_{i} \Stop_{j}} 
   - D_{33}^{124,\Stop_{j} \Stop_{j} \Stop_{j} \Sbot_{i}} 
   + D_{39}^{123,\Sbot_{i} \Sbot_{i} \Sbot_{i} \Stop_{j}} 
   - D_{39}^{123,\Stop_{j} \Stop_{j} \Stop_{j} \Sbot_{i}} \\
& 
   - D_{39}^{124,\Sbot_{i} \Sbot_{i} \Sbot_{i} \Stop_{j}} 
   + D_{39}^{124,\Stop_{j} \Stop_{j} \Stop_{j} \Sbot_{i}})
\, g[\Stop_j \Sbot_i W^\pm]\, g[ H^\pm \Stop_j \Sbot_i ]
\\
T^{\Sq,\square}_3 &=
\sum_{i,j=1}^{2}
- \frac {i\, g_S^{2} }{4\, \pi^2}
(
   - 2\,D_{27}^{132,\Sbot_{i} \Sbot_{i} \Stop_{j} \Stop_{j}}
   + 2\,D_{27}^{132,\Stop_{j} \Stop_{j} \Sbot_{i} \Sbot_{i}}
   - 2\,D_{311}^{132,\Sbot_{i} \Sbot_{i} \Stop_{j} \Stop_{j}}
   + 2\,D_{311}^{132,\Stop_{j} \Stop_{j} \Sbot_{i} \Sbot_{i}}
   + C_{0}^{34,\Sbot_{i} \Stop_{j} \Sbot_{i}} \\
&
   - C_{0}^{34,\Stop_{j} \Sbot_{i} \Stop_{j}} 
   + C_{11}^{(12) 3,\Sbot_{i} \Sbot_{i} \Stop_{j}} 
   - C_{11}^{(12) 3,\Stop_{j} \Stop_{j} \Sbot_{i}} 
   - 2\,D_{27}^{123,\Sbot_{i} \Sbot_{i} d_j \Stop_{j}}
   + 2\,D_{27}^{123,\Stop_{j} \Stop_{j} \Stop_{j} \Sbot_{i}}
   - 2\,D_{311}^{123,\Sbot_{i} \Sbot_{i} \Sbot_{i} \Stop_{j}}\\
&
   + 2\,D_{311}^{123,\Stop_{j} \Stop_{j} \Stop_{j} \Sbot_{i}}
   + 2\,D_{311}^{124,\Sbot_{i} \Sbot_{i} \Sbot_{i} \Stop_{j}}
   - 2\,D_{311}^{124,\Stop_{j} \Stop_{j} \Stop_{j} \Sbot_{i}}
   - 2\,D_{313}^{124,\Sbot_{i} \Sbot_{i} \Sbot_{i} \Stop_{j}}
   + 2\,D_{313}^{124,\Stop_{j} \Stop_{j} \Stop_{j} \Sbot_{i}}
)\\
&
\, g[\Stop_j \Sbot_i W^\pm]\, g[ H^\pm \Stop_j \Sbot_i ]
\\
T^{\Sq,\square}_4 &=  
\sum_{i,j=1}^{2}
- \frac {i\, g_S^{2} }{4\, \pi^2}
(
   - 2\,D_{313}^{132,\Sbot_{i} \Sbot_{i} \Stop_{j} \Stop_{j}}
   + 2\,D_{313}^{132,\Stop_{j} \Stop_{j} \Sbot_{i} \Sbot_{i}}
   + C_{0}^{34,\Sbot_{i} \Stop_{j} \Sbot_{i}} 
   - C_{0}^{34,\Stop_{j} \Sbot_{i} \Stop_{j}} 
   + C_{11}^{(12) 3,\Sbot_{i} \Sbot_{i} \Stop_{j}} \\
&
   - C_{11}^{(12) 3,\Stop_{j} \Stop_{j} \Sbot_{i}} 
   - 2\,D_{27}^{123,\Sbot_{i} \Sbot_{i} \Sbot_{i} \Stop_{j}}
   + 2\,D_{27}^{123,\Stop_{j} \Stop_{j} \Stop_{j} \Sbot_{i}}
   - 2\,D_{312}^{123,\Sbot_{i} \Sbot_{i} \Sbot_{i} \Stop_{j}}
   + 2\,D_{312}^{123,\Stop_{j} \Stop_{j} \Stop_{j} \Sbot_{i}} \\
&
   + 2\,D_{312}^{124,\Sbot_{i} \Sbot_{i} \Sbot_{i} \Stop_{j}}
   - 2\,D_{312}^{124,\Stop_{j} \Stop_{j} \Stop_{j} \Sbot_{i}}
   - 2\,D_{313}^{124,\Sbot_{i} \Sbot_{i} \Sbot_{i} \Stop_{j}} 
   + 2\,D_{313}^{124,\Stop_{j} \Stop_{j} \Stop_{j} \Sbot_{i}}
)\, g[\Stop_j \Sbot_i W^\pm]\, g[ H^\pm \Stop_j \Sbot_i ]
\\
T^{\Sq,\square}_5 &= 
\sum_{i,j=1}^{2}
\frac {i\, g_S^{2} }{4\, \pi^2}
(
     C_{0}^{24,\Sbot_{i} \Sbot_{i} \Stop_{j}}
   - C_{0}^{24,\Stop_{j} \Stop_{j} \Sbot_{i}} 
   - C_{11}^{(13) 2,\Sbot_{i} \Stop_{j} \Stop_{j}} 
   + C_{11}^{(13) 2,\Stop_{j} \Sbot_{i} \Sbot_{i}} 
   + 2\,D_{27}^{132,\Sbot_{i} \Sbot_{i} \Stop_{j} \Stop_{j}} \\
&
   - 2\,D_{27}^{132,\Stop_{j} \Stop_{j} \Sbot_{i} \Sbot_{i}}
   + 2\,D_{312}^{132,\Sbot_{i} \Sbot_{i} \Stop_{j} \Stop_{j}}
   - 2\,D_{312}^{132,\Stop_{j} \Stop_{j} \Sbot_{i} \Sbot_{i}}
   + 2\,D_{313}^{123,\Sbot_{i} \Sbot_{i} \Sbot_{i} \Stop_{j}}
   - 2\,D_{313}^{123,\Stop_{j} \Stop_{j} \Stop_{j} \Sbot_{i}}\\
&
   + 2\,D_{313}^{124,\Sbot_{i} \Sbot_{i} \Sbot_{i} \Stop_{j}}
   - 2\,D_{313}^{124,\Stop_{j} \Stop_{j} \Stop_{j} \Sbot_{i}}
)\, g[\Stop_j \Sbot_i W^\pm]\, g[ H^\pm \Stop_j \Sbot_i ]
\\
T^{\Sq,\square}_6 &= 
\sum_{i,j=1}^{2}
\frac {i\, g_S^{2} }{4\, \pi^2}
(
2\,D_{312}^{132,\Sbot_{i} \Sbot_{i} \Stop_{j} \Stop_{j}}
   - 2\,D_{312}^{132,\Stop_{j} \Stop_{j} \Sbot_{i} \Sbot_{i}}
   - C_{12}^{1 (23),\Sbot_{i} \Sbot_{i} \Stop_{j}} 
   + C_{12}^{1 (23),\Stop_{j} \Stop_{j} \Sbot_{i}} 
   + 2\,D_{313}^{123,\Sbot_{i} \Sbot_{i} \Sbot_{i} \Stop_{j}}\\
&
   - 2\,D_{313}^{123,\Stop_{j} \Stop_{j} \Stop_{j} \Sbot_{i}}
   + 2\,D_{313}^{124,\Sbot_{i} \Sbot_{i} \Sbot_{i} \Stop_{j}}
   - 2\,D_{313}^{124,\Stop_{j} \Stop_{j} \Stop_{j} \Sbot_{i}}
)\, g[\Stop_j \Sbot_i W^\pm]\, g[ H^\pm \Stop_j \Sbot_i ]
\end{align*}

\subsection{Helicity factors}\label{k-factors}
The factors $K_{i\, \sigma\bar{\sigma}\lambda}$ contain all
the helicity information of the amplitude. They are
obtained by contracting kinematical Tensors with the helicity
four-vectors defined in chapter~\ref{partcs}. In the following these
factors are listed. The symbol $\lambda_T$ denotes a transverse
polarization of the $W$ boson, i.e. $\lambda_T = \pm 1$).

\begin{align*}
K_{1,\sigma\bar{\sigma}0} &=  
K_{2,\sigma\bar{\sigma}0} =
\frac{
 \left( \! \,
4\,\hat{s}\,{\it |\vec{p}_W|}^{2}\,
- (\,\hat{t} - \hat{u}\,)^{2} 
 \!  \right) 
   \left( \! \,2\,{\it |\vec{p}_W|}^{2}\,\sqrt {\hat{s}} 
+ {\it E_W}\,( \hat{u} - \hat{t}) \, \!  \right)  
}{32\,(\,\hat{s}\,{\it |\vec{p}_W|}\,{\it m_{W}}\,)} \; ,\\
K_{3,\sigma\bar{\sigma}0} &= 
{\displaystyle \frac {1}{8\,{\it m_{W}}}} 
\left( \! \, - \,
2\,{\it |\vec{p}_W|}\,\sqrt {\hat{s}} + 
{\displaystyle \frac {{\it E_W}\,(
\,\hat{t} - \hat{u}\,)}{{\it |\vec{p}_W|}}}\, \!  \right) \,{\it 
(\sigma\,\bar{\sigma} + 1)}
\; ,\\
K_{4,\sigma\bar{\sigma}0} &= K_{3,\sigma\bar{\sigma}0} 
        (\hat{t} \leftrightarrow \hat{u}) 
\; ,\\
K_{5,\sigma\bar{\sigma}0} &= 
K_{6,\sigma\bar{\sigma}0} =
{\displaystyle \frac { {\it E_W} }{{\it m_{W}}} }
\left( \! \,
{\displaystyle \frac {1}{8}}\,{\displaystyle \frac {(\,\hat{t} - \hat{u}
\,)^{2}}{\hat{s}\,{\it |\vec{p}_W|}}} - {\displaystyle \frac {1}{2}}\,
{\it |\vec{p}_W|}\, \!  \right) 
\; ,
\end{align*}
\begin{align*}
K_{7,\sigma\bar{\sigma}0} &= 
        -i\,(\hat{s}/2)\,\sigma\,K_{5,\sigma\bar{\sigma}0}
\; ,\\
K_{8,\sigma\bar{\sigma}0} &= 
        i\,(\hat{s}/2)\,\bar{\sigma}\,K_{5,\sigma\bar{\sigma}0}
\; ,\\
K_{9,\sigma\bar{\sigma}0} &= 0
\; ,\\
K_{10,\sigma\bar{\sigma}0} &= 
{\displaystyle \frac { i}{16\,m_{W}}}
\left( \! \,
{\displaystyle 
\frac {{\it E_W}\,(\,\hat{t} - \hat{u}\,)}{{\it |\vec{p}_W|}}} 
- 2\,\sqrt {\hat{s}}\,{\it |\vec{p}_W|}\, 
\!  \right) \,\hat{s}\,{\it 
(\sigma + \bar{\sigma})}
\; , \\
K_{11,\sigma\bar{\sigma}0} &= K_{10,\sigma\bar{\sigma}0} 
        (\hat{t} \leftrightarrow \hat{u})
\; ,\\
K_{12,\sigma\bar{\sigma}0} &= 
K_{13,\sigma\bar{\sigma}0} = - K_{8,\sigma\bar{\sigma}0}
\; ,\\
K_{14,\sigma\bar{\sigma}0} &= 
\frac {i
\, \left( \! \,2\,{\it |\vec{p}_W|}^{2}\,\sqrt {\hat{s}} 
- {\it E_W}(\hat{t} - \hat{u}) \!  \right) 
   \left( \! \,2\,\sqrt {\hat{s}}\,{\it E_W} - \hat{t} + \hat{u}\, \! 
 \right) 
}{32\,{\it |\vec{p}_W|}\,{\it m_{W}}}
\,{\it (\sigma + \bar{\sigma})} 
\;,\\
K_{15,\sigma\bar{\sigma}0} &=
\frac {i
\, \left( \! \,2\,{\it |\vec{p}_W|}^{2}\,\sqrt {\hat{s}}
+ {\it E_W} (\hat{t} - \hat{u})   \right)
   \left( \! \,2\,\sqrt {\hat{s}}\,{\it E_W} - \hat{t} + \hat{u}\, \!
 \right)
}{32\,{\it |\vec{p}_W|}\,{\it m_{W}}}
\,{\it (\sigma + \bar{\sigma})}
\; ,
\end{align*}
\begin{align*}
  K_{16,\sigma\bar{\sigma}0} &=  
        K_{7,\sigma\bar{\sigma}0}\, m_W^2/(E_W \sqrt{\hat{s}})\; ,
& K_{17,\sigma\bar{\sigma}0}&= 
        K_{8,\sigma\bar{\sigma}0}\, m_W^2/(E_W \sqrt{\hat{s}})\; ,\\
  K_{18,\sigma\bar{\sigma}0} &= (2/\hat{s})\,K_{10,\sigma\bar{\sigma}0}\; ,
& K_{19,\sigma\bar{\sigma}0} &=  
        - K_{15,\sigma\bar{\sigma}0} (\hat{t} \leftrightarrow \hat{u})\; ,\\
  K_{20,\sigma\bar{\sigma}0} &=
        - K_{14,\sigma\bar{\sigma}0} (\hat{t} \leftrightarrow \hat{u})\; ,
& K_{21,\sigma\bar{\sigma}0} &= - K_{16,\sigma\bar{\sigma}0}\; , \\
  K_{22,\sigma\bar{\sigma}0} &= - K_{17,\sigma\bar{\sigma}0} \; ,
& K_{23,\sigma\bar{\sigma}0} &= 
        - K_{18,\sigma\bar{\sigma}0} (\hat{t} \leftrightarrow \hat{u}) \; ,\\
K_{24,\sigma\bar{\sigma}0} &=  
- \frac {i\, {\it m_{W}}\,(\,\hat{t} - \hat{u}\,) }{4\,{\it |\vec{p}_W|}\,\sqrt{\hat{s}}}\,
\,{\it (\sigma + \bar{\sigma})} \; .
\end{align*}
In the following an additional shorthand is used:
\begin{align*}
 p_t^2 & = \frac{\hat{t} \hat{u} - m_W^2 \mhpm^2}{\hat{s}} \; .
\end{align*}
\begin{align*}
K_{1,\sigma\bar{\sigma}\lambda_T} &= 
        - K_{2,\sigma\bar{\sigma}\lambda_T} =
{\displaystyle \frac {i}{32}}\, 
\left( 
\! 4\,\sqrt {\hat{s}}\,{\it |\vec{p}_W|
} - {\displaystyle \frac {\,(\,
\hat{t} - \hat{u}\,)^{2}}{{\it |\vec{p}_W|}\,\sqrt {\hat{s}}}}\, \!  \right) \,
 p_t \,\sqrt {2} \,{\it \lambda_T}
\; ,\\
K_{3,\sigma\bar{\sigma}\lambda_T} &=  
- K_{4,\sigma\bar{\sigma}\lambda_T} =
- \,{\displaystyle \frac {i}{8}}\,
{\displaystyle \frac {\sqrt {2}\, p_t \,\sqrt {\hat{s}}\,
(\,{\it \sigma}\,{\it \bar{\sigma}} + 1\,)\,{\it \lambda_T}}{{\it |\vec{p}_W|}}}
\; ,\\
K_{5,\sigma\bar{\sigma}\lambda_T} &= 
- {\displaystyle \frac {i}{8}}\, 
\left( 
\! \, {\displaystyle \frac {(\,\hat{t} - \hat{u}\,)
        \,{\it \lambda_T}}{\sqrt {\hat{s}}\,{\it |\vec{p}_W|}}} 
+ 2\,{\it \sigma}\, \!  
\right) \,
 p_t\,\sqrt {2}
\; ,\\
 K_{6,\sigma\bar{\sigma}\lambda_T} &= 
K_{5,\sigma\bar{\sigma}\lambda_T} (\sigma \to -\bar{\sigma}) \; ,\\
K_{7,\sigma\bar{\sigma}\lambda_T} &= 
{\displaystyle \frac {1}{2}}\, 
\left( 
\! \,{\displaystyle \frac {1}{4}}\,\hat{s}\,{\it \sigma}\,{\it \bar{\sigma}}
 - {\displaystyle \frac {1}{8}}\,{\displaystyle \frac {\sqrt {\hat{s}
}\,(\,\hat{t} - \hat{u}\,)\,{\it \sigma}\,{\it \lambda_T}}{{\it |\vec{p}_W|}}}\, \! 
 \right) 
\, p_t \,\sqrt {2}
\; ,\\
 K_{8,\sigma\bar{\sigma}\lambda_T} &= 
        K_{7,\sigma\bar{\sigma}\lambda_T} 
(\sigma \to -\bar{\sigma},\,\bar{\sigma} \to -\sigma)
\; ,\\
K_{9,\sigma\bar{\sigma}\lambda_T} &= 
{\displaystyle \frac {1}{8}}\,\sqrt {2}
\,\hat{s}\, p_t \,(\,{\it \sigma}\,{\it \bar{\sigma}} + 1\,)
\; ,\\
K_{10,\sigma\bar{\sigma}\lambda_T} &= 
        -K_{11,\sigma\bar{\sigma}\lambda_T} =
        i\,(\hat{s}/2)\,K_{3,\sigma\bar{\sigma}\lambda_T}
\; ,\\
K_{12,\sigma\bar{\sigma}\lambda_T} &= 
        i\,(\hat{s}/2)\,\sigma\,K_{5,\sigma\bar{\sigma}\lambda_T}
\; ,\\
 K_{13,\sigma\bar{\sigma}\lambda_T} & =
        K_{12,\sigma\bar{\sigma}\lambda_T}
        (\sigma \to -\bar{\sigma}) \; ,
\end{align*}
\begin{align*}
K_{14,\sigma\bar{\sigma}\lambda_T} &= 
        -K_{15,\sigma\bar{\sigma}\lambda_T} =
   - \,{\displaystyle \frac {1}{32}}\,{\displaystyle \frac {
 \left( \! 
\,2\,\sqrt {\hat{s}}\,{\it E_W} - \hat{t} + \hat{u}\, \! 
 \right) 
\,\sqrt {\hat{s}}\,
\, p_t \,\sqrt {2}\,
{\it (\sigma + \bar{\sigma})}\,{\it \lambda_T}}{{\it |\vec{p}_W|}}}
\; ,\\
K_{16,\sigma\bar{\sigma}\lambda_T} &= 
{\displaystyle \frac {1}{16}}
 \left( {\vrule height0.80em width0em depth0.80em} \right. \! \! 
   \left( \! 
- {\displaystyle \frac {{\it E_W}\,(\,\hat{t} - \hat{u}\,)}
        {{\it |\vec{p}_W|}}} 
+ 2\,{\it |\vec{p}_W|}\,\sqrt {\hat{s}}\, 
\! \right) 
\,{\it \sigma}\,{\it \lambda_T}
+ ( \hat{t} - \hat{u} )
- 2\,\sqrt {\hat{s}}\,{\it E_W} 
\! \! \left. {\vrule 
height0.80em width0em depth0.80em} \right)  
   p_t \,\sqrt {2} \; ,
\\
K_{17,\sigma\bar{\sigma}\lambda_T} &= 
        K_{16,\sigma\bar{\sigma}\lambda_T}
        (\sigma \to -\bar{\sigma}) \; , \\
K_{18,\sigma\bar{\sigma}\lambda_T} &= 
        K_{23,\sigma\bar{\sigma}\lambda_T} =
{\displaystyle \frac {1}{8}}\,
{\displaystyle \frac { p_t\,\sqrt {\hat{s}}\,\sqrt {2}\, 
        {\it (\sigma + \bar{\sigma})}\,{\it \lambda_T} }{{\it |\vec{p}_W|}}} 
\; ,\\
K_{19,\sigma\bar{\sigma}\lambda_T} &= 
        - K_{20,\sigma\bar{\sigma}\lambda_T} = 
        K_{15,\sigma\bar{\sigma}\lambda_T} (\hat{t} \leftrightarrow \hat{u})
\; ,\\
K_{21,\sigma\bar{\sigma}\lambda_T} &= 
{\displaystyle \frac {1}{16}}
 \left( {\vrule height0.80em width0em depth0.80em} \right. \! \! 
   \left( \! 
\,{\displaystyle \frac {{\it E_W}\,(\,\hat{t} - \hat{u}\,)}
        {{\it |\vec{p}_W|}}} 
+ 2\,{\it |\vec{p}_W|}\,\sqrt {\hat{s}}\, 
\! \right) 
\,{\it \sigma}\,{\it \lambda_T}
+ ( \hat{t} - \hat{u} )
+ 2\,\sqrt {\hat{s}}\,{\it E_W} 
\! \! \left. {\vrule 
height0.80em width0em depth0.80em} \right)  
   p_t\,\sqrt {2}
\; ,\\
K_{22,\sigma\bar{\sigma}\lambda_T} &= 
        K_{21,\sigma\bar{\sigma}\lambda_T}
        (\sigma \to -\bar{\sigma})
\; ,\\
K_{24,\sigma\bar{\sigma}\lambda_T} &=  
        - \sqrt {2}\,(E_W/\sqrt{\hat{s}})\,
        K_{18,\sigma\bar{\sigma}\lambda_T}
\; .
\end{align*}

\section{MSSM couplings}\label{couplings}

In the following all couplings of the third generation 
quarks and squarks to MSSM Higgs particles, which  
are relevant to the process, are collected. 
The factor $g_2 = e/s_w$ denotes the $SU(2)$ coupling constant of the 
weak interaction, $s_w = \sin\theta_w$, $c_w = \cos\theta_w$
and $t_w = \tan\theta_w$ with the weak mixing angle $\theta_w$.
The scalar and pseudoscalar couplings in the Higgs--fermion interactions
are distiguished by adding subscripts 's' and 'p' to the coupling symbols.
Analogously, the subscripts 'v' and 'a' distiguish the vector and 
axial vector coupling of quarks to the $W$ boson. 

\subsection{Quark couplings to Higgs bosons}

\subsubsection*{Neutral Higgs bosons} 

\begin{align}
g_s[ H^0 t t ] & =
        -g_2\:\frac{m_t}{2 m_W}\:\frac{\sin\alpha}{\sin\beta} 
\; ,
&  
g_s[ H^0 b b ] & =
        -g_2\:\frac{m_b}{2m_W}\:\frac{\cos\alpha}{\cos\beta}
\; ,\\
g_s[ h^0 t t ] & =
        -g_2\:\frac{m_t}{2m_W}\:\frac{\cos\alpha}{\sin\beta}
\; ,
&
g_s[ h^0 b b ] & =
        +g_2\:\frac{m_b}{2m_W}\:\frac{\sin\alpha}{\cos\beta}
\; ,\\
g_p[ H^0 t t ] & = g_p[ h^0 t t ] = 0
\; ,
&
g_p[ H^0 b b ] & = g_p[ h^0 b b ] = 0
\; ,\\
g_p[ A^0 t t ] & = i g_2\:\frac{m_t}{2 m_W} \cot\beta
\; ,
&
g_p[ A^0 b b ] & = ig_2\:\frac{m_b}{2 m_W} \tan\beta
\; ,\\
g_s[ A^0 t t ] & = g_s[ A^0 b b ] = 0
\; .
\end{align}

\subsubsection*{Charged Higgs bosons}

\begin{align}
g_s[H^+\:\text{out}, t\:\text{in}, b\:\text{out}] & =
        g_2\,\frac{m_b\tan\beta + m_t\cot\beta}{2\sqrt{2} m_W} =
        g_s[H^-\:\text{out}, b\:\text{in}, t\:\text{out}] 
\; ,\\
g_p[H^+\:\text{out}, t\:\text{in}, b\:\text{out}] & =
        -g_2\,\frac{m_b\tan\beta - m_t\cot\beta}{2\sqrt{2} m_W} =
        -g_p[H^-\:\text{out}, b\:\text{in}, t\:\text{out}]
\; .
\end{align}

\subsection{Squark couplings to Higgs bosons}

\subsubsection*{Neutral Higgs bosons}

\begin{multline}
g[ H^0 \Stop_1 \Stop_1 ] = g_2\, 
\Big[  \frac{m_t}{2 m_W \sin\beta} 
        \Big( (A_t \sin\alpha - \mu\cos\alpha) \sin 2\theta_{\Stop}
                -2 m_t \sin\alpha 
        \Big) \\
      + \frac{m_Z \cos(\alpha+\beta)}{6 c_w} 
        \Big( (5 - 8 c_w^2) \cos^2\theta_{\Stop} -4 s_w^2
        \Big)   
\Big]
\; ,
\end{multline}

\begin{multline}
g[ H^0 \Stop_2 \Stop_2 ] = g_2\, 
\Big[  \frac{m_t}{2 m_W \sin\beta} 
        \Big( - (A_t \sin\alpha - \mu\cos\alpha) \sin 2\theta_{\Stop}
                -2 m_t \sin\alpha 
        \Big) \\
      + \frac{m_Z \cos(\alpha+\beta)}{6 c_w} 
        \Big( - (5 - 8 c_w^2) \cos^2\theta_{\Stop} + (1 -4 s_w^2)
        \Big)   
\Big]
\; ,
\end{multline}

\begin{multline}
g[ H^0 \Sbot_1 \Sbot_1 ] = g_2\, 
\Big[  \frac{m_b}{2 m_W \cos\beta} 
        \Big( (A_b \cos\alpha -\mu\sin\alpha ) \sin 2\theta_{\Sbot}
                -2 m_b \cos\alpha
        \Big) \\
      + \frac{m_Z \cos(\alpha+\beta)}{6 c_w} 
        \Big( (4 c_w^2 - 1 ) \cos^2\theta_{\Sbot} + 2 s_w^2
        \Big)   
\Big]
\; ,
\end{multline}

\begin{multline}
g[ H^0 \Sbot_2 \Sbot_2 ] = g_2\, 
\Big[  \frac{m_b}{2 m_W \cos\beta} 
        \Big( - (A_b \cos\alpha -\mu\sin\alpha ) \sin 2\theta_{\Sbot}
                -2 m_b \cos\alpha
        \Big) \\
      + \frac{m_Z \cos(\alpha+\beta)}{6 c_w} 
        \Big( - (4 c_w^2 - 1 ) \cos^2\theta_{\Sbot} + (1 + 2 c_w^2)
        \Big)   
\Big]
\; ,
\end{multline}

\begin{align}
g[ h^0 \Sq_i \Sq_i ] & = g[H^0,\Sq_i,\Sq_i] \:
\big( \sin\alpha \to \cos\alpha\, ,\: \cos\alpha \to -\sin\alpha 
\big)
\; .
\end{align}

\subsubsection*{Charged Higgs bosons}

\begin{multline}
g[ H^\pm \Stop_1 \Sbot_1 ]  =
\frac{ { g_2} }{2 \sqrt{2} { m_W} } \Big[
+ 2 m_t m_b ( \tan\beta + \cot\beta ) \: \sin\theta_{\Sbot} \sin\theta_{\Stop}\\
        \sin 2 \beta \Big( m_b^2 (1 + \tan^2\beta )
                         + m_t^2 (1 + \cot^2\beta ) - 2 m_W^2 \Big) 
        \:\cos\theta_{\Sbot} \cos\theta_{\Stop}\\
- 2 m_b ( \mu + A_b \tan\beta ) \: \sin\theta_{\Sbot} \cos\theta_{\Stop}
- 2 m_t ( \mu + A_t \cot\beta ) \: \cos\theta_{\Sbot} \sin\theta_{\Stop}
\Big]
\; ,
\end{multline}

\begin{multline}
g[ H^\pm \Stop_1 \Sbot_2 ] = 
\frac{ { g_2} }{2 \sqrt{2} { m_W} } \Big[
- 2 m_t m_b ( \tan\beta + \cot\beta ) \: \cos\theta_{\Sbot} \sin\theta_{\Stop}\\
+\sin 2 \beta \Big( m_b^2 (1 + \tan^2\beta ) 
                         + m_t^2 (1 + \cot^2\beta ) - 2 m_W^2 \Big) 
        \:\sin\theta_{\Sbot} \cos\theta_{\Stop}\\
+  2 m_b ( \mu + A_b \tan\beta ) \: \cos\theta_{\Sbot} \cos\theta_{\Stop}
- 2 m_t ( \mu + A_t \cot\beta ) \: \sin\theta_{\Sbot} \sin\theta_{\Stop}
\Big]
\; ,
\end{multline}

\begin{multline}
g[ H^\pm \Stop_2 \Sbot_1 ] = 
\frac{ { g_2} }{2 \sqrt{2} { m_W} } \Big[
- 2 m_t m_b ( \tan\beta + \cot\beta ) \: \sin\theta_{\Sbot} \cos\theta_{\Stop}\\
+\sin 2 \beta \Big( m_b^2 (1 + \tan^2\beta ) 
                         + m_t^2 (1 + \cot^2\beta ) - 2 m_W^2 \Big) 
        \:\cos\theta_{\Sbot} \sin\theta_{\Stop}\\
+ 2 m_b ( \mu + A_b \tan\beta ) \: \sin\theta_{\Sbot} \sin\theta_{\Stop}
 2 m_t ( \mu + A_t \cot\beta ) \: \cos\theta_{\Sbot} \cos\theta_{\Stop}
\Big]
\; ,
\end{multline}

\begin{multline}
g[ H^\pm \Stop_2 \Sbot_2 ] =
\frac{ { g_2} }{2 \sqrt{2} { m_W} } \Big[
 2 m_t m_b ( \tan\beta + \cot\beta ) \: \cos\theta_{\Sbot} \cos\theta_{\Stop}\\
+\sin 2 \beta \Big( m_b^2 (1 + \tan^2\beta ) 
                         + m_t^2 (1 + \cot^2\beta ) - 2 m_W^2 \Big) 
        \:\sin\theta_{\Sbot} \sin\theta_{\Stop}\\
+ 2 m_b ( \mu + A_b \tan\beta ) \: \cos\theta_{\Sbot} \sin\theta_{\Stop}  
+ 2 m_t ( \mu + A_t \cot\beta ) \: \sin\theta_{\Sbot} \cos\theta_{\Stop}
\Big]
\; .
\end{multline}

\subsection{Couplings to Gauge bosons}
\subsubsection*{Quarks and Squarks}
The indices $\alpha,\beta,a,b$ are colour indices; $i$ and $j$ denote
squark mass-eigenstates.
\begin{align}
g_v[ q^\alpha q^\beta g^a ] & = g_S \frac{\lambda^a_{\alpha\beta}}{2}& g_a[ q q g ] & = 0 
\; ,\\
g[\Sq^\alpha \Sq^\beta g^a ] & = -g_S \frac{\lambda^a_{\alpha\beta}}{2} 
\; ,\\
g[\Sq^\alpha \Sq^\beta g^a g^b] & = g_S^2 
        \big\{ \frac{\lambda^a}{2}, \frac{\lambda^b}{2} \big\}_{\alpha\beta}
\; ,\\
g_v[t^\beta b^\alpha W^\pm] & = \frac{g_2}{2\sqrt{2}} \delta_{\alpha\beta}
        & g_a[t^\beta b^\alpha W^\pm] & = \frac{g_2}{2\sqrt{2}} \delta_{\alpha\beta}
\; ,\\
g[\Stop_i \Sbot_j W^\pm] & = - \frac{g_2}{\sqrt{2}} R_{ij} 
\; ,\\
g[\Stop_i^\beta \Sbot_j^\alpha g^a W^\pm] & = \sqrt{2} g_2 g_S R_{ij} 
        \frac{\lambda^a_{\alpha\beta}}{2}
\; .
\end{align}
The $R_{ij}$ are the following functions of the Squark mixing angles:
\begin{align}
\left(
\begin{array}[2]{cc}
R_{11} & R_{12}\\
R_{21} & R_{22} 
\end{array}
\right)
= 
\left(
\begin{array}[2]{cc}
\cos\theta_{\Stop}\cos\theta_{\Sbot} & \sin\theta_{\Stop}\cos\theta_{\Sbot}\\
\cos\theta_{\Stop}\sin\theta_{\Sbot} & \sin\theta_{\Stop}\sin\theta_{\Sbot}
\end{array} 
\right)
\; .
\end{align}   
\subsubsection*{Higgs bosons}
\begin{align}
g[ h^0 H^\pm W^\mp] & = \frac{g_2}{2} \sin(\beta - \alpha)
\; ,\\
g[ H^0 H^\pm W^\mp] & =-\frac{g_2}{2} \cos(\beta - \alpha)
\; ,\\
g[ A^0 H^\pm W^\mp] & =-i \frac{g_2}{2}
\; .
\end{align}


\section{Feynman graphs}\label{feynman-graphs}

\begin{fmffile}{gghw-graphs}

Feynman Graphs with opposite direction of charge flow are not 
depicted.

\subsection{Quark graphs}\label{q-graphs}
\vspace*{.5cm}

\begin{center}

\begin{tabular}[2]{cc}
\begin{picture} (60,30)(-10,-5)
\begin{fmfgraph*}(40,28)
\fmfleft{i1,i2}
\fmfright{o1,o2}
\fmflabel{$W^-$}{o1}
\fmflabel{$H^+$}{o2}
\fmf{gluon}{i1,v1}
\fmf{gluon}{i2,v2}
\fmf{fermion,tension=.55,label=$t$,l.side=left}{v1,v2}
\fmf{fermion,tension=.55,label=$t$,l.side=left}{v2,v3}
\fmf{fermion,tension=.55,label=$b$,l.side=left}{v3,v4}
\fmf{fermion,tension=.55,label=$t$,l.side=left}{v4,v1}
\fmf{zigzag}{o1,v4}
\fmf{scalar}{v3,o2}
\fmfdot{v1,v2,v3,v4}
\end{fmfgraph*}
\end{picture}
&
\begin{picture} (60,30)(-10,-5)
\begin{fmfgraph*}(40,28)
\fmfleft{i1,i2}
\fmfright{o1,o2}
\fmf{gluon}{i1,v1}
\fmf{gluon}{i2,v2}
\fmf{fermion,tension=.55,label=$b$,l.side=left}{v1,v2}
\fmf{fermion,tension=.55,label=$b$,l.side=left}{v2,v3}
\fmf{fermion,tension=.55,label=$t$,l.side=right}{v3,v4}
\fmf{fermion,tension=.55,label=$b$,l.side=left}{v4,v1}
\fmf{phantom}{o1,v4}
\fmf{phantom}{v3,o2}
\fmffreeze
\fmf{scalar}{v4,o2}
\fmf{zigzag}{v3,o1}
\fmfdot{v1,v2,v3,v4}
\end{fmfgraph*}
\end{picture}
\end{tabular}

\begin{tabular}[2]{cc}
\begin{picture} (60,30)(-10,-1)
\begin{fmfgraph*}(40,28)
\fmfleft{i1,i2}
\fmfright{o1,o2}
\fmf{gluon}{i2,v2}
\fmf{phantom}{i1,v1}
\fmf{fermion,tension=.55,label=$t$,l.side=left}{v1,v2}
\fmf{fermion,tension=.55,label=$t$,l.side=left}{v2,v3}
\fmf{fermion,tension=.55,label=$b$,l.side=left}{v3,v4}
\fmf{fermion,tension=.55,label=$b$,l.side=right}{v4,v1}
\fmf{scalar}{v3,o2}
\fmf{phantom}{o1,v4}
\fmffreeze
\fmf{gluon}{i1,v4}
\fmf{zigzag}{v1,o1}
\fmfdot{v1,v2,v3,v4}
\end{fmfgraph*}
\end{picture}
&
\begin{picture} (60,30)(-5,-1)
\begin{fmfgraph*}(50,28)
\fmfleft{i1,i2}
\fmfright{o1,o2}
\fmf{gluon}{i1,v1}
\fmf{gluon}{i2,v2}
\fmf{fermion,tension=.3,label=$t/b$,l.side=left}{v1,v2}
\fmf{fermion,tension=.3,label=$t/b$,l.side=left}{v2,v3}
\fmf{fermion,tension=.3,label=$t/b$,l.side=left}{v3,v1}
\fmf{dashes,tension=.4,label=$h^0/H^0/A^0$,l.side=left}{v3,v4}
\fmf{scalar}{v4,o2}
\fmf{zigzag}{o1,v4}
\fmfdot{v1,v2,v3,v4}
\end{fmfgraph*}
\end{picture}
\end{tabular}
\end{center}

\subsection{Squark-Graphs}\label{sq-graphs}
\vspace*{.5cm}

\begin{center}

\begin{tabular}[2]{cc}
\begin{picture} (60,30)(0,-1)
\begin{fmfgraph*}(60,28)
\fmfleft{i1,i2}
\fmfright{o1,o2}
\fmflabel{$g$}{i1}
\fmflabel{$g$}{i2}
\fmflabel{$W^-$}{o1}
\fmflabel{$H^+$}{o2}
\fmf{gluon}{i1,v1}
\fmf{gluon}{i2,v1}
\fmf{scalar,left,tension=.5,label=$\Stop_i/\Sbot_i$,l.side=left}{v1,v2}
\fmf{scalar,left,tension=.5,label=$\Stop_i/\Sbot_i$,l.side=left}{v2,v1}
\fmf{dashes,tension=.8,label=$h^0/H^0$,l.side=left}{v2,v3}
\fmf{scalar}{v3,o2}
\fmf{zigzag}{o1,v3}
\fmfdot{v1,v2,v3}
\end{fmfgraph*}
\end{picture}
&
\begin{picture} (60,30)(-5,-1)
\begin{fmfgraph*}(50,28)
\fmfleft{i1,i2}
\fmfright{o1,o2}
\fmf{gluon}{i1,v1}
\fmf{gluon}{i2,v2}
\fmf{scalar,tension=.3,label=$\Stop_i/\Sbot_i$,l.side=left}{v1,v2}
\fmf{scalar,tension=.2,label=$\Stop_i/\Sbot_i$,l.side=left}{v2,v3}
\fmf{scalar,tension=.2,label=$\Stop_i/\Sbot_i$,l.side=left}{v3,v1}
\fmf{dashes,tension=.3,label=$h^0/H^0$,l.side=left}{v3,v4}
\fmf{scalar}{v4,o2}
\fmf{zigzag}{o1,v4}
\fmfdot{v1,v2,v3,v4}
\end{fmfgraph*}
\end{picture}
\end{tabular}

\begin{tabular}[2]{cc}
\begin{picture} (60,30)(0,-1)
\begin{fmfgraph*}(40,28)
\fmfleft{i1,i2}
\fmfright{o1,o2}
\fmf{gluon}{i1,v1}
\fmf{gluon}{i2,v2}
\fmf{scalar}{v3,o2}
\fmf{scalar,tension=.5,label=$\Stop_j$,l.side=left}{v1,v2}
\fmf{scalar,tension=.3,label=$\Stop_j$,l.side=left}{v2,v3}
\fmf{scalar,tension=.5,label=$\Sbot_i$,l.side=left}{v3,v1}
\fmf{zigzag}{o1,v1}
\fmfdot{v1,v2,v3}
\end{fmfgraph*}
\end{picture}
&
\begin{picture} (60,30)(0,-1)
\begin{fmfgraph*}(40,28)
\fmfleft{i1,i2}
\fmfright{o1,o2}
\fmf{gluon}{i1,v1}
\fmf{gluon}{i2,v2}
\fmf{scalar}{v3,o2}
\fmf{scalar,tension=.5,label=$\Sbot_j$,l.side=right}{v2,v1}
\fmf{scalar,tension=.3,label=$\Sbot_j$,l.side=right}{v3,v2}
\fmf{scalar,tension=.5,label=$\Stop_i$,l.side=right}{v1,v3}
\fmf{zigzag}{o1,v1}
\fmfdot{v1,v2,v3}
\end{fmfgraph*}
\end{picture}
\end{tabular}

\begin{tabular}[2]{cc}
\begin{picture} (60,30)(0,-1)
\begin{fmfgraph*}(40,28)
\fmfleft{i1,i2}
\fmfright{o1,o2}
\fmf{gluon}{i1,v1}
\fmf{gluon}{i2,v1}
\fmf{scalar,tension=.5,label=$\Sbot_i$,l.side=right}{v1,v2}
\fmf{scalar,tension=.2,label=$\Stop_i$,l.side=right}{v2,v3}
\fmf{scalar,tension=.5,label=$\Sbot_i$,l.side=right}{v3,v1}
\fmf{zigzag}{o1,v2}
\fmf{scalar}{v3,o2}
\fmfdot{v1,v2,v3}
\end{fmfgraph*}
\end{picture}
&
\begin{picture} (60,30)(0,-1)
\begin{fmfgraph*}(40,28)
\fmfleft{i1,i2}
\fmfright{o1,o2}
\fmf{gluon}{i1,v1}
\fmf{gluon}{i2,v1}
\fmf{scalar,tension=.5,label=$\Stop_i$,l.side=left}{v1,v3}
\fmf{scalar,tension=.2,label=$\Sbot_i$,l.side=left}{v3,v2}
\fmf{scalar,tension=.5,label=$\Stop_i$,l.side=left}{v2,v1}
\fmf{zigzag}{o1,v2}
\fmf{scalar}{v3,o2}
\fmfdot{v1,v2,v3}
\end{fmfgraph*}
\end{picture}
\end{tabular}

\begin{tabular}[2]{cc}
\begin{picture} (60,30)(0,-1)
\begin{fmfgraph*}(40,28)
\fmfleft{i1,i2}
\fmfright{o1,o2}
\fmf{gluon}{i1,v1}
\fmf{gluon}{i2,v2}
\fmf{scalar,tension=.55,label=$\Sbot_j$,l.side=right}{v1,v4}
\fmf{scalar,tension=.55,label=$\Stop_i$,l.side=right}{v4,v3}
\fmf{scalar,tension=.55,label=$\Sbot_j$,l.side=right}{v3,v2}
\fmf{scalar,tension=.55,label=$\Sbot_j$,l.side=right}{v2,v1}
\fmf{zigzag}{o1,v4}
\fmf{scalar}{v3,o2}
\fmfdot{v1,v2,v3,v4}
\end{fmfgraph*}
\end{picture}
&
\begin{picture} (60,30)(0,-1)
\begin{fmfgraph*}(40,28)
\fmfleft{i1,i2}
\fmfright{o1,o2}
\fmf{gluon}{i1,v1}
\fmf{gluon}{i2,v2}
\fmf{scalar,tension=.55,label=$\Stop_i$,l.side=left}{v1,v2}
\fmf{scalar,tension=.55,label=$\Stop_i$,l.side=left}{v2,v3}
\fmf{scalar,tension=.55,label=$\Sbot_j$,l.side=left}{v3,v4}
\fmf{scalar,tension=.55,label=$\Stop_i$,l.side=left}{v4,v1}
\fmf{zigzag}{o1,v4}
\fmf{scalar}{v3,o2}
\fmfdot{v1,v2,v3,v4}
\end{fmfgraph*}
\end{picture}
\end{tabular}

\begin{tabular}[1]{c}
\begin{picture} (40,30)(0,-1)
\begin{fmfgraph*}(40,28)
\fmfleft{i1,i2}
\fmfright{o1,o2}
\fmf{gluon}{i2,v2}
\fmf{phantom}{i1,v1}
\fmf{scalar,tension=.55,label=$\Stop_i$,l.side=left}{v1,v4}
\fmf{scalar,tension=.55,label=$\Stop_i$,l.side=right}{v4,v3}
\fmf{scalar,tension=.55,label=$\Sbot_j$,l.side=right}{v3,v2}
\fmf{scalar,tension=.55,label=$\Sbot_j$,l.side=right}{v2,v1}
\fmf{scalar}{v3,o2}
\fmf{phantom}{v4,o1}
\fmffreeze
\fmf{gluon}{i1,v4}
\fmf{zigzag}{o1,v1}
\fmfdot{v1,v2,v3,v4}
\end{fmfgraph*}
\end{picture}
\end{tabular}
\end{center}

\end{fmffile}


\newpage

\section*{Figure Captions}
\begin{description}
\item[Figure 1] 
Partonic cross section evaluated using
        all Feynman graphs (solid lines), only quark loop graphs
        (dashed lines) and only squark loop graphs (dotted lines)
        for a sequence of charged-Higgs masses
        ($100,210,360,450 \,\gev$) and $\tan\beta = 1.5$.
        The parameter set 1 (see table \ref{mixing-cases})
        is used in the squark sector.
\item[Figure 2]
Hadronic cross section for $H^+ W^-$
        production via gluon fusion versus the charged-Higgs
        mass $\mhpm$ for two values of $\tan\beta$ (1.5,6).
        The three squark scenarios without mixing (A, B, C in Table
        \ref{nomixing-cases}) are compared with the case of decoupling
        squarks (solid lines).
\item[Figure 3]
Hadronic cross section for $H^+ W^-$
        production via gluon fusion versus $\tan\beta$ for
        three values of the charged-Higgs mass $\mhpm$ (100,300,1000 $\gev$).
        The three squark cases without mixing (A, B, C in Table
        \ref{nomixing-cases}) are compared
        with the case of decoupling squarks (solid lines).
\item[Figure 4]
Hadronic cross section for $H^+ W^-$
        production via gluon fusion versus the charged-Higgs
        mass $\mhpm$ for two values of $\tan\beta$ (1.5,6).
        The three squark scenarios with large mixing  (1, 2, 3 in Table
        \ref{mixing-cases}) are compared with the case of decoupling
        squarks (solid lines).
\item[Figure 5]
Hadronic cross section for $H^+ W^-$
        production via gluon fusion versus $\tan\beta$ for
        three values of the charged-Higgs mass $\mhpm$ (100,360,1000 $\gev$).
        The squark case 1 (thick lines) is compared to the case of
        decoupling squarks (thin lines).
\item[Figure 6]
Hadronic cross section for $H^+ W^-$
        production via gluon fusion versus $\tan\beta$ for
        three values of the charged-Higgs mass $\mhpm$ (100,410,1000 $\gev$).
        The squark case 2 (thick lines) is compared to the case of
        decoupling squarks (thin lines).
\item[Figure 7]
Hadronic cross section for $H^+ W^-$
        production via gluon fusion versus $\tan\beta$ for
        three values of the charged-Higgs mass $\mhpm$ (100,470,1000 $\gev$).
        The squark case 3 (thick lines) is compared to the case of
        decoupling squarks (thin lines).
\end{description}


\end{document}